\documentclass[aps,prx,twocolumn,floats,showpacs,superscriptaddress]{revtex4-1}
\usepackage{amsthm,amssymb,amsmath}
\usepackage[top=20mm, bottom=20mm, left=15mm, right=15mm]{geometry}
\usepackage{graphicx,epsfig}
\usepackage{graphicx}
\usepackage{float}
\usepackage{times}
\usepackage{graphics,dcolumn,bm,float}
\usepackage{amssymb,amsmath,rotate,color}
\usepackage[pagebackref=false,colorlinks,linkcolor=blue,citecolor=blue,urlcolor=blue]{hyperref}
\usepackage{xcolor}
\usepackage{ulem}

\begin{document}
\unitlength 1.5 cm
\newcommand{\be}{\begin{equation}}
\newcommand{\ee}{\end{equation}}
\newcommand{\bea}{\begin{eqnarray}}
\newcommand{\eea}{\end{eqnarray}}
\newcommand{\nn}{\nonumber}
\newcommand{\hlt}[1]{\textcolor{red}{#1}}
\newcommand{\up}{\uparrow}
\newcommand{\down}{\downarrow}

\newcommand{\bl}[1]{\textcolor{blue}{#1}}
\newcommand{\re}[1]{\textcolor{red}{#1}}
\newcommand{\gr}[1]{\textcolor{green}{#1}}

\title{Phase transitions in the binary-alloy Hubbard model:\\ insight from strong-coupling perturbation theory}
\author{Elaheh Adibi} 
\affiliation{Department of Physics, Sharif University of Technology, Tehran 11155-9161, Iran}
\author{Alireza Habibi} 
\affiliation{Department of Physics, Sharif University of Technology, Tehran 11155-9161, Iran}
\affiliation{School of Physics, Institute for Research in Fundamental Sciences (IPM), Tehran 19395-5531, Iran}
\author{S. A. Jafari}
\email{jafari@physics.sharif.edu}
\affiliation{Department of Physics, Sharif University of Technology, Tehran 11155-9161, Iran}
\affiliation{Center of excellence for Complex Systems and Condensed Matter (CSCM), Sharif University of Technology, Tehran 14588-89694, Iran}

\begin{abstract} 
In the binary-alloy with composition $A_xB_{1-x}$ of two atoms with ionic energy scales $\pm \Delta$, 
an apparent Anderson insulator (AI) is obtained as a result of randomness in the position of atoms. Using our recently developed 
technique that combines the local self-energy from strong-coupling perturbation theory with the transfer matrix method, we are able to 
address the problem of adding a Hubbard $U$ to the binary alloy problem for millions of lattice sites on the honeycomb lattice. 
By adding the Hubbard interaction $U$, the resulting AI phase will become metallic which in our formulation can be clearly 
attributed to the screening of disorder by Hubbard $U$. Upon further increase in $U$, again the AI phase emerges which can be understood 
in terms of the suppressed charge fluctuations due to residual Hubbard interaction of which the randomness takes advantage and localizes 
the quasi-particles of the metallic phase. The ultimate destiny of the system at very large $U$ is to become a Mott insulator (MI). 
We construct the phase diagram of this model in the plane of $(U,\Delta)$ for various compositions $x$. 
\end{abstract} 
\maketitle
\section{Introduction}
Coulomb interaction between electrons plays a significant role in the electronic structure of the real materials. 
The metal to insulator transition arising from correlations, i.e., 
Mott transition,  have been the subject of intensive research in the past decades~\cite{sorella1992,jafari2009,Wu10,He12,Hassan13,Otsuka16,Adibi16,MottBook}. 
On the reverse direction, a considerable amount of new theoretical work is devoted to the correlation driven {\it insulator to metal} phase transition. In other words,
the question is whether the correlations can drive an insulating phase to a metallic phase or not? A well-studied example is to start from a band insulating (BI) 
phase obtained from a periodic external potential. Adding Hubbard interactions to the periodic potential gives rise to the so-called ionic Hubbard 
model~\cite{Garg06,Paris-IHM07,Bouadim07,Craco08,Ebrahimkhas12,Adibi16}. 
These studies show that by increasing interaction strength, the half-filled ionic Hubbard model has two transition points.
The first transition is from BI to metal. And of course, by further increase of the interaction strength, the second transition point involves a transition 
between the metal and  MI.

 The other route to the interaction driven transformation of insulator into metal is provided by the disordered systems. 
 The disorder is introduced through random on-site energies which is distributed according to some probability distribution. 
 Choosing a uniform distribution by self-consistent Hartree-Fock calculations and quantum Monte Carlo (QMC) method it was 
 found that two-dimensional system undergoes an AI to the metallic phase transition~\cite{Denteneer99,Denteneer03,Chakraborty07,Heidarian04}.
In these studies the lattice sizes used in calculations were small. Therefore, the AI to metal transition can not be determined with a good accuracy.
 In contrast, the metallic phase induced by correlations was not reported in the phase diagram of the Refs.~\cite{Henseler08,Atkinson08}. 
 Also, in Ref.~\cite{Habibi18} we ruled out the possibility of any metallic state between the AI and MI  for the uniformly distributed Anderson disorder. 
 
In the present work, instead of Anderson disorder, we consider binary-alloy disorder $A_xB_{1-x}$ that is composed of two different atoms $A$ and $B$ which are randomly distributed
on the lattice. Once the positions of $A$ and $B$ in a random configuration are chosen, each atom $A$ ($B$) is assigned an on-site potential $+\Delta$ ($-\Delta$).
In all sites, a Hubbard term $U$ operates. The resulting model is the so-called binary-alloy Hubbard model.  
The spatially ordered limit of the binary-alloy system on a bipartite lattice corresponds to the ionic Hubbard model. 
Refs.~\cite{Balzer05} and~\cite{Lombardo06} studied the binary-alloy Hubbard model by using mean field theory on the three-dimensional lattice
and dynamical mean field theory (DMFT) on the Bethe lattice, respectively. They obtained that at half-filling, i.e.,  having on average one electron at every lattice site, two metal-insulator transitions occur by increasing the Hubbard interaction strength. The first transition point corresponds to the transition from an uncorrelated BI to a metal, and the second transition is from a metallic phase to MI. 
Employing DMFT, Byczuk~\textit{et al.} found that a new metal to Mott-type insulator transition occurs because of the interplay between band splitting by binary-alloy disorder and correlation effects~\cite{Byczuk03}. The numerical renormalization group study at zero temperature has shown that the system becomes a MI at strong interactions at incommensurate densities $n=x$ or $n=1+x$~\cite{Byczuk04}.  
In Ref.~\cite{Paris07}, Paris~\textit{et al.} investigated the phase diagram of the Hubbard model with the binary-alloy disorder on the square lattice. 
They found the MI behavior away from half-filling in agreement with previously mentioned studies. Combining QMC simulations 
and exact diagonalization, they were also able to treat disorder better than the earlier mean field and DMFT studies to explore the possibility of AI phase.

The correlation effects in the honeycomb lattice have been widely investigated which result in a number of exotic phenomena in both theory 
and experiment such as the correlated electrons in the graphene~\cite{Castro09,Kotov12} and Silicene~\cite{Verri07,Vogt12,Houssa15,BaskaranSilicene} 
as well as topological Mott insulator~\cite{Raghu08}. 
Combination of disorder and correlation gives rise to other interesting phenomena such as the formation of AI phases and
possible transitions to metallic behavior driven by interactions. 

We have recently been able to integrate the strong-coupling perturbation theory -- which can analytically address the Mott transition --
with very efficient numerical methods of disorder problems~\cite{Habibi18}. The local
nature of strong-coupling self-energy allows an efficient embedding into the transfer matrix method,
which is essentially free of size limitations and therefore enables us to perform a careful and reliable finite size scaling.
Generally speaking, the strong-coupling perturbation theory is based on an exact treatment of the atomic limit, followed by switching on the inter-site hopping~\cite{Senechal98,Senechal2000}. Our present combination of the two powerful tools of Mott and Anderson physics,  
allows us to treat correlation and disorder on equal footing. Computation of the Green's function for 
very large lattice is the important advantage of this method. By utilizing the Green's function, 
the Kernel polynomial method (KPM) enables us to compute the density of state (DOS) in real space for a disordered and interacting system with millions of lattice sites. KPM as a highly efficient numerical method provides a high precision determination of DOS without an explicit diagonalization of the Hamiltonian \cite{KPM,SotaRKPM,Habibi13}. 
On the other hand, the transfer matrix method which can numerically let the particles propagate 
on the lattice up to essentially any desirable length scale, can be nicely combined with the
local self-energy obtained from strong-coupling perturbation theory to immediately give us the
localization length of the wave function in a background that includes, not only on-site (random) energies,
but also the appropriately trained local self-energies. 
This method enables a reliable finite size scaling which in turn furnishes valuable insight into 
the localization properties of strongly correlated and disordered system.

In this paper, we set out to study the binary-alloy Hubbard model with our method on the honeycomb lattice.
Our key finding is that competition between local binary disorder and electronic correlations leads to a metallic phase. 
This is in contrast to Anderson disorder (non-binary alloy) where a metallic phase between MI and AI is ruled out~\cite{Habibi18}. 
Besides, we find the AI phase as the ground state of the system in a specific region of parameters whereas some methods 
such as mean field and DMFT used in Refs.~\cite{Balzer05,Lombardo06} failed to identify this phase. 
Our results are backed by a careful finite size scaling which due to the exponential growth of the Hilbert space,
is not possible in methods that treat the Hubbard part numerically. 

The rest of this paper is structured as follows.  In Sec.~\ref{Model and Method} we introduce the alloy-disordered Hubbard model followed by brief 
review of the strong-coupling perturbation approach. After that, we present the results of our calculations in 
Sec.~\ref{results}. Finally, in Sec.~\ref{conclusion}, we end up with some concluding remarks. The Appendices provide the self-energy formulae along with 
brief description of KPM and transfer matrix method used in present work.

\section{Strong-coupling perturbation theory\label{Model and Method}}
In this section we extend the strong-coupling perturbation formalism of strongly interacting systems 
to include the disorder as well. The canonical model of disordered interacting system is the Anderson-Hubbard model, 
which is given by the following Hamiltonian,
\bea
&&H=H_0+H_1,\label{Hamiltonian}\\
&&H_0=U \sum_{i} n_{i\uparrow} n_{i\downarrow}-\mu\ \sum_{i,\sigma}\ n_{i\sigma}+\sum_{i,\sigma}\ \epsilon_i\ n_{i\sigma},\label{unperturbed-H}\nn\\
&&H_1=\sum_{ij,\sigma}\ V_{ij}\ (c_{i\sigma}^{\dagger}\ c_{j\sigma}+H.c.)\label{perturbation},\nn
\eea
where $H_0$ accounts for interaction and disorder energy, and $H_1$ for kinetic energy. Also $c_{i\sigma}^{\dagger}$ ($c_{i\sigma}$) is the fermionic creation (annihilation) operator of the particle with spin $\sigma=\pm1/2$ on the lattice site $i$, $n_{i\sigma}=c_{i\sigma}^{\dagger}\ c_{i\sigma}$ is the local electron number operator, $V_{ij}$ is the hopping matrix element between sites $i$ and $j$, and $U>0$ is the on-site Coulomb repulsion. The chemical potential $\mu$  determines the average density of the system. 
The local energies $\epsilon_i$ are independent random variables. In the following, we assume a binary probability distribution for $\epsilon_i$, i.e.,
\begin{align}
   P(\epsilon_i)=x\ \delta(\epsilon_i-\Delta) + (1-x)\ \delta(\epsilon_i+\Delta).
\end{align}
Here $\Delta$ parameterizes the binary-alloy disorder strength.  $x$ and $1-x$ are the concentrations of the two components of the 
alloy ions with energies $\Delta$ and $-\Delta$, respectively. Additionally, $x=0$ and $x=1$ correspond to non-disordered system with shifted on-site energy $\pm\Delta$.

To be self-contained, 
in the following, we briefly describe the strong-coupling perturbation theory. 
Considering $H_0$ and $H_1$ in Hamiltonian (\ref{Hamiltonian}) as the unperturbed and perturbed Hamiltonian respectively, the partition function at temperature $T=1/\beta$ in the path-integral formalism is written as,
\bea
Z&=&\int [d\gamma^{\star} d\gamma]\ \exp\bigg[ -\int_{0}^{\beta}d\tau\bigg\lbrace \sum_{i\sigma} \gamma^{\star}_{i\sigma}(\tau)\ \partial_\tau\ \gamma_{i\sigma}(\tau)\nn\\
&+&H_0(\gamma^{\star}_{i\sigma}(\tau),\gamma_{i\sigma}(\tau))+\sum_{ij\sigma}\gamma^{\star}_{i\sigma}(\tau)\ V_{ij}\ \gamma_{j\sigma}(\tau) \bigg\rbrace\bigg], 
\eea
where $\gamma$ and $\gamma^{\star}$ denote the Grassmann fields in the imaginary time $\tau$. 

As can be seen, the the unperturbed part of the Hamiltonian is not quadratic. Therefore, the ordinary Wick theorem can not be employed to construct a diagrammatic expansion for the correlation functions. To circumvent this, one starts with the following Hubbard-Stratonovich transformation~\cite{Senechal2000}, 
\bea
&&\int [d\psi^\star d\psi] \exp\bigg[\int_{0}^{\beta} d\tau \sum_{i\sigma} \bigg\lbrace\sum_j \psi^\star_{i\sigma}(\tau) (V^{-1})_{ij} \psi_{j\sigma}(\tau)\nn\\
&&\qquad\qquad+\psi^\star_{i\sigma}(\tau)\gamma_{i\sigma}(\tau)+\gamma^\star_{i\sigma}(\tau)\psi_{i\sigma}(\tau)\bigg\rbrace\bigg]\nn\\
&&=\det(V^{-1})\exp\bigg[-\int_{0}^{\beta} d\tau \sum_{ij\sigma} \gamma^\star_{i\sigma}(\tau)\ V_{ij}\ \gamma_{j\sigma}(\tau) \bigg],
\eea
In this equation $\psi_{i\sigma}(\tau)$ and $\psi^\star_{i\sigma}(\tau)$ are the auxiliary Grassmann fields. 
So, by means of this transformation, we can rewrite the partition function up to a normalization factor as,
\bea
Z=\int[d\psi^\star d\psi] \exp\bigg[-\bigg\lbrace S_0[\psi^\star,\psi]+\sum_{R=1}^{\infty} S^R_{int}[\psi^\star,\psi]\bigg\rbrace \bigg].\label{rewritten-Z}
\eea
In the above partition function,  $S_0[\psi^\star,\psi]$ is the free (unperturbed) auxiliary fermion action which is 
determined by the inverse of the hopping matrix of original fermions, 
\bea
S_0[\psi^\star,\psi]=-\int_{0}^{\beta}\ d\tau \sum_{ij\sigma} \psi^\star_{i\sigma}(\tau)\ (V^{-1})_{ij}\ \psi_{j\sigma}(\tau),
\eea
and $S^{R}_{int}[\psi^\star,\psi]$  includes an infinite number of interaction terms,
\bea
 &&S^{R}_{int}[\psi^\star,\psi]=\frac{-1}{(R!)^2}\sum_{i}\sum_{\lbrace\sigma_l\sigma^{\prime}_l\rbrace} \int_{0}^{\beta}\ \prod_{l=1}^{R}\ d\tau_l d\tau'_l\nn\\ &\times &\psi_{i\sigma_1}^{\star}(\tau_1)\ldots\psi_{i\sigma_R}^{\star}(\tau_R)\psi_{i\sigma'_R}(\tau'_R)\ldots\psi_{i\sigma'_1}(\tau'_1)\nn\\
&\times &\bigg\langle \gamma_{i\sigma_1}(\tau_1)\ldots\gamma_{i\sigma_R}(\tau_R)\gamma^{\star}_{i\sigma'_R}(\tau'_R)\ldots\gamma^{\star}_{i\sigma'_1}(\tau'_1)\bigg\rangle_{0,c}.\label{interaction-Z}
\eea
The $\langle \gamma_{i\sigma_1}(\tau_1)\ldots\gamma_{i\sigma_R}(\tau_R)\gamma^{\star}_{i\sigma'_R}(\tau'_R)\ldots\gamma^{\star}_{i\sigma'_1}(\tau'_1)\rangle_{0,c}$ 
represents connected correlation function of the original fermions, which now determines the interaction vertices of the dual theory. 
In the partition function of the auxiliary fermions (\ref{rewritten-Z}), the unperturbed part is quadratic. Hence, the Wick theorem can be applied to 
take the interaction term (\ref{interaction-Z}) perturbatively into account and calculate the self-energy 
of the auxiliary fermion ($\Gamma$). Finally, the Green's function of the original fermions is expressed by,
\bea
G=(\Gamma^{-1}-V)^{-1}\label{green'sfunction}.
\eea
The diagrammatic details of the strong-coupling approach can be found in Ref.~\onlinecite{Senechal2000}.

\begin{figure}[tb]
    \includegraphics[width=2.4cm]{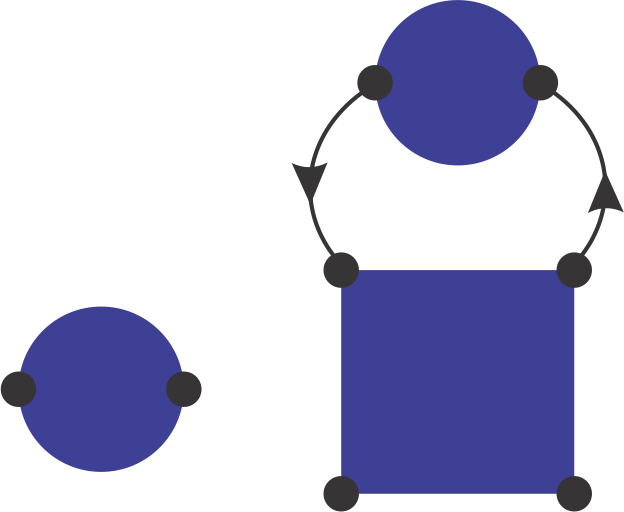}
    \caption{Diagrams contributing to the self-energy of the auxiliary fermions up to order $t^2$ where solid lines indicate free propagator 
    $V$ of auxiliary fermions and vertices represent connected correlation functions. Circles are one-particle connected correlation functions and square refers to two-particle connected correlation function.}
    \label{diagram}
\end{figure}

\section{Results and discussion\label{results}}
In this section, we present our results obtained by strong-coupling expansion for the Hubbard model with the binary-alloy disorder on the honeycomb lattice. 
We assume $V_{ij}=-t$ if $i,j$ are nearest neighbor sites and  zero otherwise. Throughout the paper, $t = 1$ sets the unit of the energy. 
In the absence of disorder, setting $\mu=U/2$ corresponds to half-filling. However, for the system affected by the disorder, this chemical potential does not necessarily correspond to the half-filling. To determine the half-filling in the presence of disorder, we numerically solve the 
implicit equation, $\overline{n(\mu)}=\int_{-\infty}^\mu \overline{\rho(\varepsilon)}\ d\varepsilon=1/2$,
where bar denotes averaging over realizations of the disorder. The numerical cost involved here is due to 
the fact that $\rho$ implicitly depends on $\mu$. So $\mu$ has to be self-consistently
satisfy the above equation. 

In this paper, in order to study the disordered interacting electrons on the honeycomb lattice the perturbative treatment up to order  $t^2$ is considered.  
We can compute the self-energy of the auxiliary fermions up to second order as depicted in the Feynman diagrams in Fig.~\ref{diagram}.  
Therefore, the self-energy  for each spin is expressed as,
\bea
\Gamma_{ij}(i\omega)=\Gamma_{ij}^{(0)}(i\omega)+\Gamma_{ij}^{(2)}(i\omega),
\eea
where the $\Gamma_{ij}^{(0)}(i\omega)$ and $\Gamma_{ij}^{(2)}(i\omega)$ are the zeroth and second order self-energy for the auxiliary fermions, respectively. 
The Matsubara frequencies are $i\omega=i(2n+1)\pi T$. The algebraic expressions for the $\Gamma_{ij}^{(0)}(i\omega)$ and $\Gamma_{ij}^{(2)}(i\omega)$ at arbitrary temperature and fixed chemical potential are presented in Appendix \ref{self-energy}.
Once the self-energy of the auxiliary fermions is obtained, one can calculate the Green's function by Eq.~\eqref{green'sfunction}.

\subsection{Fixed chemical potential $\mu=U/2$} 
 
To investigate the physics of strongly correlated and binary-disordered electrons on the honeycomb lattice, first we study the DOS which is given by,
\bea
   \rho(\omega)=\frac{1}{N}\sum_{n=1}^{N}\ \delta(\omega-E_n),
\eea 
where $N$ is the number of lattice sites and $E_n$ denotes the eigenvalues of the Hamiltonian. The DOS can be determined very 
efficiently by the KPM~\cite{KPM,SotaRKPM,Habibi13} which is described in Appendix~\ref{kpm}. 
 \begin{figure}[b]
     \centering
     \includegraphics[width=1\linewidth]{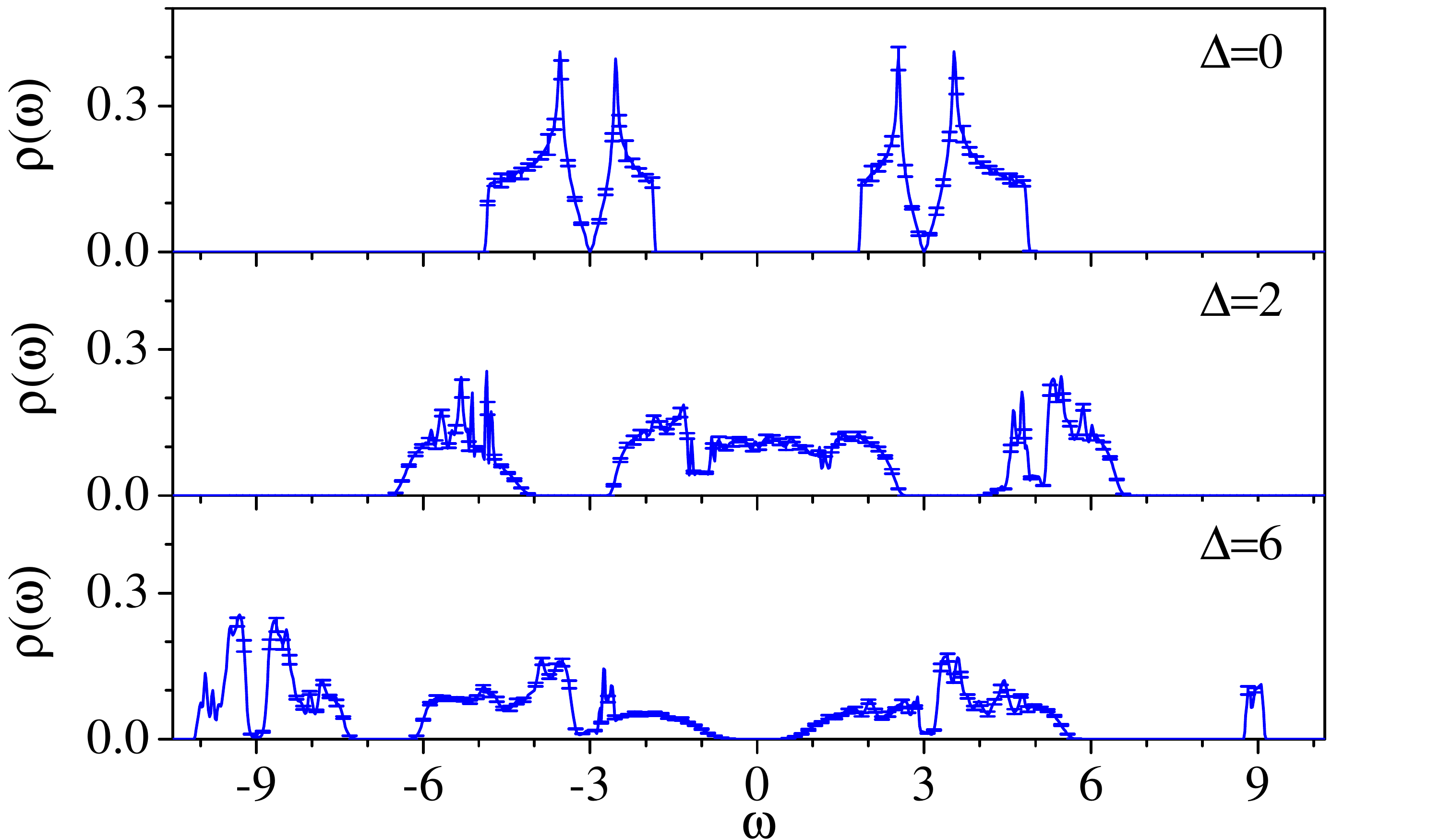}
     \caption{(Color online) Disorder-averaged DOS obtained by KPM for different disorder strengths $\Delta$ at 
     concentration $x=0.5$,  $\mu=U/2$ and interaction strength $U=6$. The system size is $500\times 500$.
     }
     \label{fig:dos-u6}
 \end{figure}
In Fig.~\ref{fig:dos-u6}  DOS has been plotted for several disorder strengths at interaction $U=6$, $\mu=U/2$ and zero temperature 
for disorder concentration $x=0.5$ and system with $500\times 500$ lattice sites. In the absence of disorder, i.e., $\Delta=0$, an 
insulating phase corresponding to MI can be observed in agreement with Refs.~\cite{Adibi16,Otsuka16}.
For large enough interaction strengths we expect the DOS of the clean system to have two main sub-bands around $-U/2$ and $U/2$ 
which corresponds to upper and lower Hubbard sub-bands. Moreover, this band splitting is proportional to the interaction strength $U$, which is a hallmark of Mott-Hubbard bands. 

Turning the disorder on, we can discuss how DOS is affected by the disorder. After introducing disorder to the system, DOS will consist of four branches which correspond to Hubbard sub-bands plus an additional band splitting within each Hubbard sub-band that is originated from ionic energies. This result is in agreement with Ref.~\cite{Adibi16} which reports similar features in the ionic Hubbard model.
Generally, disorder broadens each branch of DOS spectrum and reduces Mott gap as reported in e.g. Ref.~\cite{Habibi18}. 
The Mott gap is robust against weak disorder. By increasing the disorder strength, the level repulsion pushes the two disorder-split sub-bands of each
Hubbard band towards each other which eventually results in gap closing. So as can be seen at $\Delta=2$ in Fig.~\ref{fig:dos-u6} the 
gap is completely filled. However, here DOS can not determine the conductive nature of the gapless states emerged at the Fermi level (i.e. at $\omega=0$)
due to competition between the interaction and randomness. Upon further increase of the disorder strength for $\Delta \approx 5$ a gap reopens in the spectrum. 
Therefore, as shown in Fig.~\ref{fig:dos-u6} the DOS exhibits an energy gap in the spectrum at interaction strength $\Delta=6$. 

The DOS in Fig.~\ref{fig:dos-u6} for $\Delta=2$ reveals that for this disorder strength the $\mu=U/2$ specifies the half-filling. 
In contrast for $\Delta=6$, setting $\mu=U/2$ no longer specifies Fermi energy at half-filling. 
Furthermore, closure of the Mott gap in the presence of disorder indicates that disorder shifts the 
Mott transition to larger values of interaction strengths. This behavior is  captured by dual fermion approach~\cite{Haase17} and 
DMFT~\cite{Byczuk05} as well.

   \begin{figure}[b]
     \centering
     \includegraphics[width=0.9\linewidth]{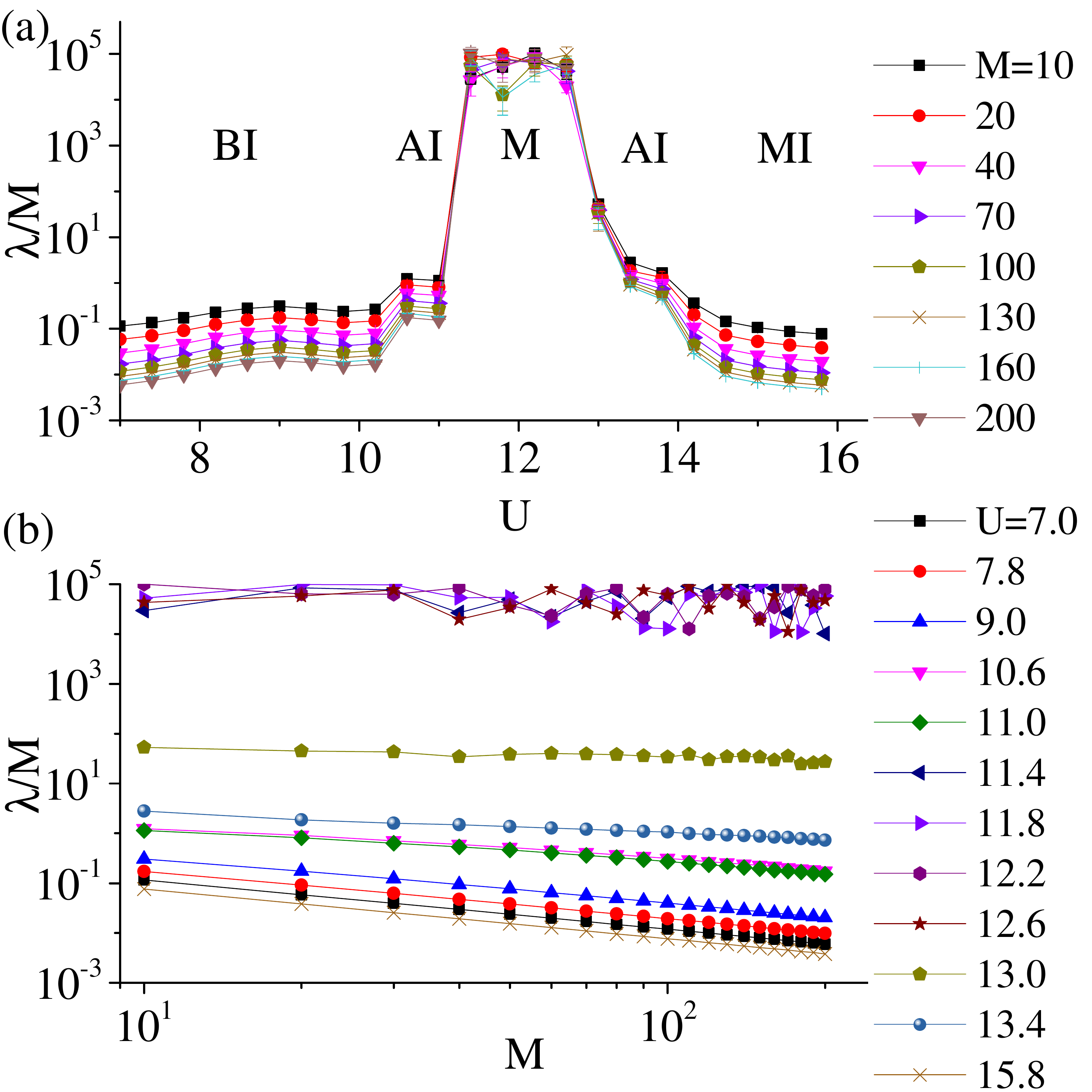}
     \caption{(Color online) The localization length normalized to the width in ribbon geometry for honeycomb lattice at Fermi energy ($\omega=0$) and fixed chemical potential $\mu=U/2$ as a function of (a) interaction strength $U$ for various ribbon widths $M$ indicated in the legend, and (b) the ribbon width $M$ for various Hubbard interaction $U$. In both cases the disorder strength $\Delta = 6$ and concentration $x=0.5$ is considered. }
     \label{fig:lambda-fixmu}
 \end{figure}

As can be seen in the DOS profile, disorder suppresses the Mott gap and gives rise to gapless state first. 
Given that the system is disordered, the question will be, whether the resulting gapless state is a metal or not?
To characterize the nature of the gapless phases arising from the interplay of binary-alloy disorder and interaction, 
we employ the transfer matrix method which is briefly explained in Appendix \ref{TMM}. The transfer matrix method as a 
powerful technique allows us to determine the localization length of the disordered systems with a large number of atoms. 
This method enables a reliable finite size scaling which in turn will allow us to determine that the gapless states around the Fermi energy have the extended or localized properties. For finite size scaling, in Fig.~\ref{fig:lambda-fixmu} 
we plot the localization length normalized to the ribbon width, $\lambda/M$, in 
ribbon geometry  of the honeycomb lattice at fixed chemical potential $\mu=U/2$, disorder strength $\Delta=6$ and 
composition $x=0.5$  for lattice with length $L=100000$. In panel (a) the 
interaction dependence of normalized localization length for various ribbon widths $M$ is displayed.
As can be seen for a given width, at weak interaction strengths the localization length takes very small values which indicates the insulating behavior. This insulating state at large disorder strength $\Delta=6$ corresponds to BI (gapped) which is perturbatively connected to the non-interacting case $U=0$, and continue to weak interactions $U$. 
Increasing the Hubbard interaction the normalized localization length increases up to the onset of a jump around $U\approx 10.4$.  
This is followed by another bigger jump at $U\approx 11.2$. 
According to the DOS, the region between the above two values of $U$ is gapless, but the localization length values are still
small to give a conducting phase. The finite size scaling in the lower panel (b) confirms this observation.
Therefore it has to correspond to localized AI phase. Upon further increase of interaction strength, a pronounced plateau 
appears in the localization length. In this region we still have a gapless phase, but with much larger localization lengths.
The finite size scaling in panel (b) confirms that it corresponds to the metallic phase. Hence, the competition between on-site Hubbard correlation 
and binary-ally form of randomness gives rise to a metallic ground state which passes through an AI phase. 
Increasing the interaction strength, the localization length shows the localized behavior again before entering the (gapped) MI phase at strong $U$.

Let us explain in more detail the finite size scaling which is necessary to corroborate the phases identified in panel (a) of Fig.~\ref{fig:lambda-fixmu}. In panel (b) of this figure 
we plot the normalized localization length as a function of ribbon width $M$ for different interaction strengths. 
In general, in Mott/band insulating (gapped) and Anderson localized (gapless) phases the localization length decreases with increasing the width. 
Albeit the localization length for Mott and band insulators is lower compared to neighboring Anderson localized phases. 
This difference has to do with the fact that Mott and band insulators are gapped, while the Anderson insulator is gapless. 
To interpret the curves, note that increasing behavior the localization length versus width, $M$, means that at infinite lattice we have infinite localization 
length corresponding to extended state and hence the system will be a metal. Such behavior can be seen in panel (b)
for $U=11.4,11.8,12.2,12.6$ which confirms that the localization length plateau in panel (a) does indeed correspond to a conducting phase.
So from this point let us call it the metallic plateau. 
Fitting an appropriate function for different value of disorder, the normalized localization length indicates the relation $\lambda/M \approx M^{-f(U,\Delta,M)}$,
where $f>0$ ($f\le0$) implies that system is localized (extended)~\cite{Habibi18}.
 
 The essential lesson to be learned from Fig.~\ref{fig:lambda-fixmu} is that the effect of correlations on localization is to initially 
suppress the localization. This can be interpreted as the screening of the disorder~\cite{Henseler08,Atkinson08}
by repulsion $U$ until it reaches a maximum value at the metallic plateau. 
On the right side of the metallic plateau, there are two players: One is a very strong Hubbard $U$ which in
the conducting background of the metallic plateau region with a residual Hubbard $U$ will be able to generate
the substantial super-exchange interaction that wants to drive the system towards the ultimate Mott phase.
In the absence of the random $\pm \Delta$ alloy potential, the $U_{\rm res.}/t$ would be the only player and one would have the Mott phase. However, 
the randomness in the binary-alloy potential takes advantage of the fact that charge fluctuations are suppressed by residual Hubbard, $U_{\rm res}$, 
and will be able to localize them giving
rise to AI phase easily. This AI phase is followed by Mott-Hubbard splitting of the spectral weight, ending the system in MI phase. 
In this way, the transition from a disordered metal to a Mott phase in the right side of the localization plateau is preceded by an AI phase.

    \begin{figure}[tb]
    \centering
    \includegraphics[width=0.9\linewidth]{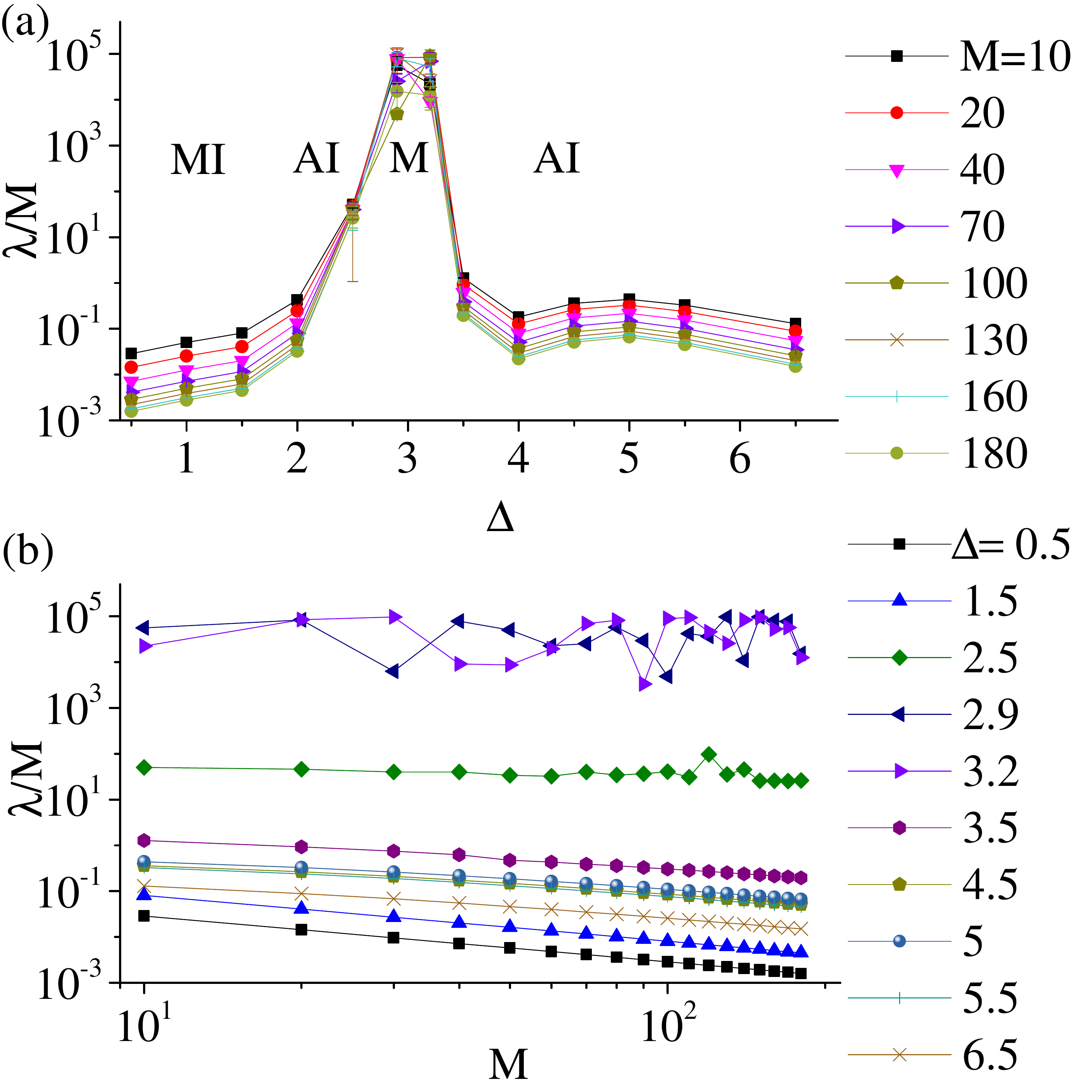}
    \caption{ (Color online) 
        The  localization length normalized to the width, in ribbon geometry  calculated by transfer matrix method at zero energy for half-filled 
        binary alloy Hubbard model with a fixed interaction strength $U=6$ and alloy composition $x=0.5$, as a function of (a) disorder 
        strength $\Delta$ for different ribbon widths $M$ indicated in the legend, and (b) the ribbon width $M$ for various $\Delta$. 
    }
    \label{fig:lambda}
\end{figure}

\subsection{Half-filling}
Up to this point, we have been focused on fixed chemical potential $\mu=U/2$, but as can be inferred from Fig.~\ref{fig:dos-u6}, 
the $\mu=U/2$ does not necessarily correspond to the half-filling. In other words, for some $\Delta_c$, when $\Delta>\Delta_c$,
the condition $\mu=U/2$ does not specify the half-filling.  Considering the half-filled case, the localization length normalized 
to width, in ribbon geometry is plotted in Fig.~\ref{fig:lambda} for a fixed interaction strength $U=6$ and alloy composition $x=0.5$. 
The normalized localization length versus disorder strength $\Delta$ for different values of the ribbon width is shown in panel (a). 
Vanishing DOS at $U=6$ and weak disorder strength means that the state at small $\Delta$ is a MI (gapped). 
Therefore, there are no states near Fermi energy which consequently leads to very small localization length. 
Increasing disorder fills in the Mott gap. 
This is because for the larger $\Delta$, on average half of the sites (since $x=0.5$) will have a large negative alloy potential
which denies the no-double-occupancy rule of the Mott phase and hence substantial charge fluctuations are created which will then fill in the Mott gap. 
The Mott gap closure results in increasing the localization length. But the states created around the Fermi level 
are not yet ready for conduction, as the density of states is small which means that we are dealing with a poor metal that can be
localized by moderate disorder in the position of the alloy potential $\pm\Delta$.
Finite size scaling in panel (b) of Fig.~\ref{fig:lambda} confirms that the state to the right of MI is AI. 
By further increase in $\Delta$, the system enters a metallic plateau where the localization length is maximal. 
Eventually the phase in the right of metallic plateau is again an AI. {\it But at half-filling, there is no BI phase in large $\Delta$}. 

The metallic plateau in Fig.~\ref{fig:lambda} can be explained
as follows: Imagine that the alloy components A and B (at $x=0.5$) are arranged in a regular
bipartite lattice. Then at half-filling, a competition between $\Delta$ and $U$ will generate a conducting phase for $\Delta\approx U/2$
within strong-coupling perturbation theory~\cite{Adibi16}. Similar picture is obtained by 
DMFT~\cite{Garg06,Ebrahimkhas12} and continuous unitary transformation (CUT)~\cite{Hafez2009CUT,ShahnazHafez}. So the metallic phase, in this
case, can be considered a descendant of the conducting phase in the above studies. Adding disorder on
top of such a conducting state naturally explains the AI to the right of metallic phase in Fig.~\ref{fig:lambda}.
The transition from the leftmost MI to AI is similar a direct transition between MI and AI in other
disordered systems~\cite{Habibi18,Haase17}. However the transition from the AI in the left of metallic plateau
to conducting phase in the plateau is unusual. Technically this happens because when $\Delta$ approaches $U/2$,
the self-energy $\Gamma$ of auxiliary fermions diverges. By Eq.~\eqref{green'sfunction} this divergence
implies the Green's function will be similar to those of free fermions. It looks like that by approaching
$\Delta\approx U/2$, the interaction and binary-alloy disorder knockout each other, and we are 
left with a conducting phase.

\begin{figure}[t]
     \centering
     \includegraphics[width=1\linewidth]{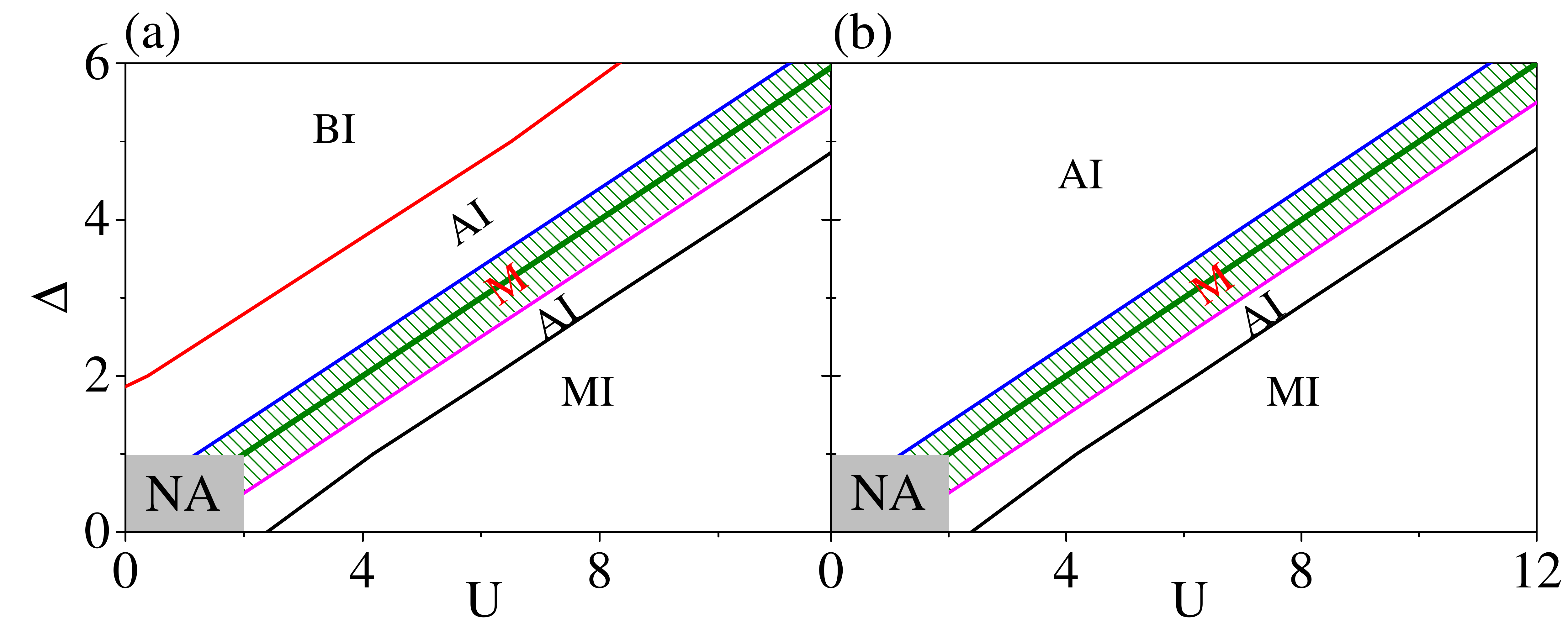}
     \caption{Ground state phase diagram of  Coulomb interaction $U$ versus  disorder strength $\Delta$ of the Hubbard model 
     with binary-alloy disorder in concentration $x=0.5$ for (a) fixed chemical potential $\mu=U/2$ and (b) half-filled case.
     The NA region corresponds to a weak $U$ where the strong-coupling perturbation expansion
     is not reliable. 
     }
     \label{fig:phasediagram}
\end{figure}

As pointed out, DOS tells us whether we have gapped state or gapless state. Then by a finite size scaling of the localization length, 
we can determine the nature of the single-particle states around the Fermi surface in the gapless state. 
Therefore, by computing the DOS and localization length for various Hubbard interaction $U$ and disorder strength $\Delta$ 
we can generate the phase diagram in $(U,\Delta)$ plane for the correlated fermions on honeycomb lattice with the binary-alloy disorder. 
The phase diagram for disorder concentration $x=0.5$ in two cases are plotted in 
Fig.~\ref{fig:phasediagram}: (a) fixed chemical potential $\mu=U/2$ and (b) half-filling.  
We must stress again that the fixed $\mu=U/2$ corresponds to half-filling only for weak enough 
$\Delta<\Delta_c$. So the Mott phase which happens in $\Delta<\Delta_c$, is identical in both panels (a) and (b). 
Since the strong-coupling expansion is reliable at strong-coupling limit, in calculating the Mott gap extracted from DOS,  
first, we compute the single particle gap for large interaction strengths at a given fixed  $\Delta$.
Then, by extrapolating the gap to zero, we can find the critical Coulomb repulsion $U_c$ for Mott transition 
for a fixed disorder strength $\Delta$~\cite{Adibi16,Sahebsara08,Senechal2000}. 

By varying Hubbard $U$ for a fixed disorder strength $\Delta$ in Fig.~\ref{fig:phasediagram}(a), we can
span phases from weakly correlated phase BI to strongly correlated MI. 
The NA region corresponds to small $U$ where the strong-coupling expansion is not applicable. 
We, therefore, discuss outside the NA region. 
As can be seen, between these two insulating phases, a metallic phase surrounded by AI phase has emerged. 
For the half-filled case as presented in Fig.~\ref{fig:phasediagram}(b), the BI phase disappears in comparison with the phase diagram of $\mu=U/2$. 
Indeed, as pointed out earlier in the limit $\Delta \gg U$, $\mu=U/2$ does not correspond to half-filling. To maintain the half-filling, the spectral weight 
must be shifted to higher energies to keep the Fermi level at $\omega=0$. 
So if the system was gapped for $\mu=U/2$, after the spectral shift, the DOS becomes nonzero at Fermi level. 
Therefore in the half-filled the case, the large $\Delta/U$ BI of panel (a) is never realized in panel (b). 

It is worth noting that the green line shaded area in both panel (a) and (b) of Fig.~\ref{fig:phasediagram}, 
represents the metallic phase is centered around the green line $U=2\Delta$~\cite{Adibi16} where $2\Delta$ denotes the energy difference 
of alloy energies. As pointed out earlier, at $\Delta=U/2$ the self-energy $\Gamma(\omega)$ diverges at Fermi level ($\omega=0$) and consequently by
Eq.~\eqref{green'sfunction}, interaction completely screens the disorder. Hence, the system does not see any disorder which results in a 
robust metallic phase which persists in both situations with fixed chemical potential, and fixed density. 
In one hand by infinitesimal deviation from $U=2\Delta$, the self-energy will start to feel the disorder in the system. 
This can be viewed as the partial screening of the disorder by interaction. On the other hand small disorder on the honeycomb lattice, 
does not localize the wave functions in the middle of the band, and the system remains metal~\cite{Heidarian04,Amini09,Gorbachevprl2009,Tatiana2014}. 
However, far away from the $U=2\Delta$, the interaction can not screen disorder, and metallic phase no longer persists.
Note that the AI phase that is sandwiched between metallic and MI phase is narrower than the other AI phase. 
The former AI takes advantage of the reduces charge fluctuations due to large $U$. But once this AI phase
is established, it will be easier for the Hubbard $U$ to stabilize a Mott phase on a AI background. The other
AI on the other hand has to compete with the ionic energy scale of a disordered BI phase in panel (a). Larger $\Delta$
means more disorder, and more ionicity which makes both BI and AI phases happy. That is why the AI phase to the left of
metallic state spans a larger region.

\subsection{Dependence on composition ratio $x$}
\begin{figure}[tb]
     \centering
     \includegraphics[width=1\linewidth]{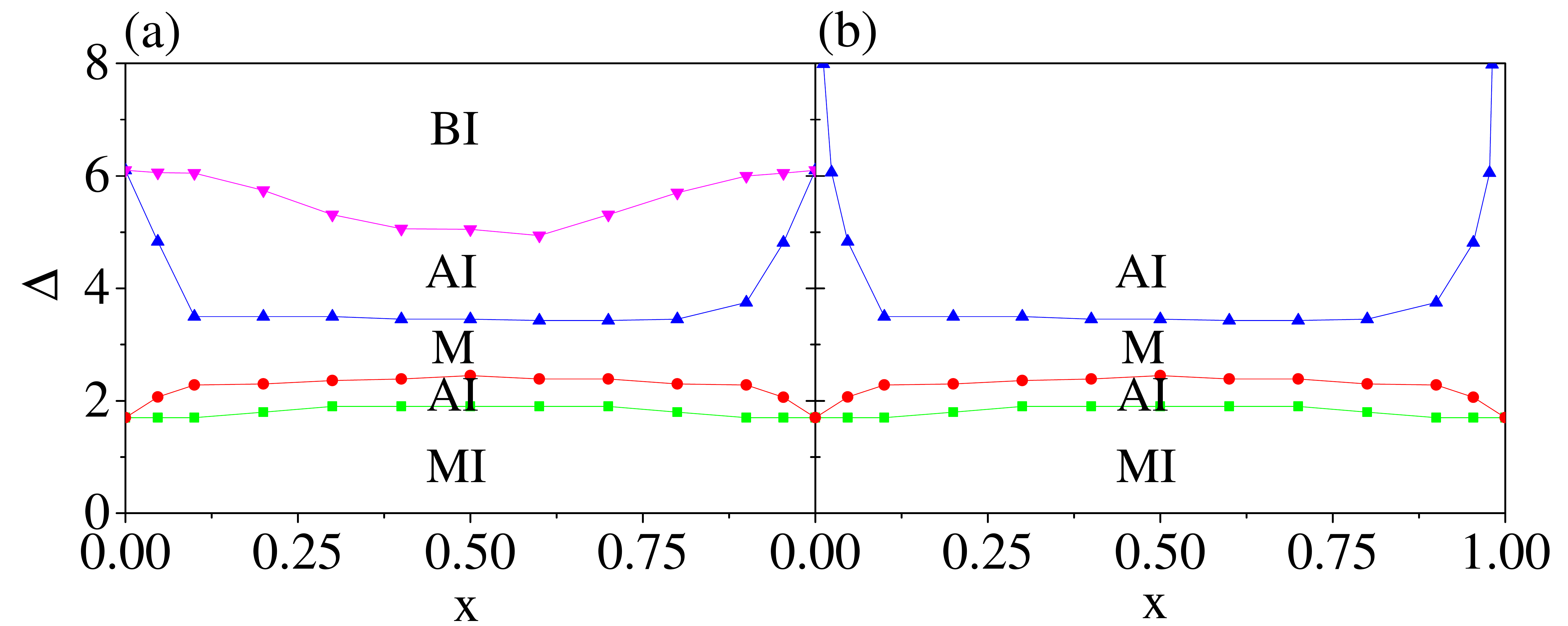}
     \caption{Phase diagram at fixed interaction strength $U=6$ as a function of  disorder concentrations $x$ for (a) $\mu=U/2$ and (b) half-filling.}
     \label{fig:x-delta}
\end{figure}

So far we have fixed the disorder concentration at $x=0.5$, the regular limit of which corresponds to the ionic Hubbard model.
This helps to (i) identify the gapped phase in weakly correlated phase at fixed chemical potential with BI,
(ii) understand the nature of the middle metallic phase. Let us see how this picture carries on for other composition ratios $x$,
which might be more relevant to realistic materials than $x=0.5$. 
So we examine other compositions by varying $x$. 
To this end, let us start by construction of the phase diagram of the Anderson-Hubbard model for various concentrations $x$ at 
fixed $U=6$ as displayed in Fig.~\ref{fig:x-delta} for (a) $\mu=U/2$ and (b) half-filling. 
Changing the concentration of atoms $A$ with energy $\Delta$ from $x=0.5$ to $x=0$, atoms $B$ with energy $-\Delta$ constitute the 
major portion of the system such that at $x=0$ only $B$ atoms exist. So the main branch of DOS belongs to atoms $B$.
Assuming $x=0$ in Fig.~\ref{fig:x-delta}(a), for a fixed on-site repulsion $U$, and fixed $\mu=U/2$ and $\Delta=0$ system is MI. 
Upon switching on the $\Delta$, the spectral weight transfers towards $\omega\approx -\Delta$. 
The rearrangement of the spectral weight is controlled by the random variations in the inverse self-energy $\Gamma^{-1}$ of the auxiliary fermions.  
The spectral weight transfer for $\Delta$ smaller than the certain limit, say $\Delta_{c1}$ (the green square border) is 
such that still the Fermi level will remain inside the Mott gap. At $\Delta=\Delta_{c1}$ the Fermi level crosses the shifted spectrum at a non-zero DOS 
and hence the Mott gap will be pushed to energies well above (below) 
the Fermi level and therefore we have basically a gapless phase up to some upper limit $\Delta<\Delta_{c2}$.  
$\Delta_{c2}$ (the pink down-triangle) is the critical strength for which all states shifted to the right of the Fermi energy 
and there are no states in Fermi level anymore. So for $\Delta>\Delta_{c2}$ system is an empty band and hence will be in the (gapped) BI phase.
The same explanation is valid for $x=1$, because in this limit system is composed of only $A$ atoms with on-site energy $\Delta$.
The difference between $x=1$ and $x=0$ cases is that the direction of the spectral weight shift is reversed. 
Since at $x=0$ and $x=1$ the system has disorder, we do not have any Anderson insulator phase. That is why by approaching
these two points the AI phase is shrunk to zero. For $x\in (0,1)$ we have AI phases with a metallic phase sandwiched between them. 

In panel (b) of Fig.~\ref{fig:x-delta} all possible phases in a fixed particle density corresponding to half-filling have been plotted. 
Half-filling is obtained by calculating the chemical potential self-consistently at every $x$ and $\Delta$. For $x=0$ and $x=1$ 
which corresponds to ordered arrangements of atoms, as explained in panel (a) by increasing $\Delta$ the system becomes MI which is 
followed by gapless phase. After closing the Mott gap, due to half-filling, Fermi level always falls in the regions with non-zero DOS
and the system remains metal. If the composition ratio $x$ deviates from $x=0$ or $x=1$, depending on the $\Delta$ magnitude we start
to obtain AI states similar to $x=0.5$.

Why in the fixed chemical potential $\mu=U/2$ case 
the large $\Delta/U$ region is a 
BI while in the half-filled case with $n=1$ no gapped (BI) state at large $\Delta/U$ is obtained?
Physically fixing the particle density (e.g., here at half-filling) means that we are dealing with a closed system 
to/from which electrons can not be added/removed. In this case, in the rightmost
AI phases of Fig.~\ref{fig:lambda}(a), electrons are localized wherever they are. By increasing $\Delta$ the alloy potential wells
become deeper and every electron stays where it is. So we are dealing with a compressible state
where an extra electron can be added to a suitable location at negligible cost. However, when the
chemical potential is maintained constant by an external gate, those regions whose alloy potential
is $-\Delta$, suck more electrons from the gate, and eventually, all the sites with negative alloy
potential are filled. Therefore for large enough $\Delta$, we expect an incompressible (gapped) state
adding electrons to which requires finite energy. That is why in the constant chemical potential
case the large $\Delta$ region is a BI while with fixed particle density, no BI follows the AI phase
by increasing $\Delta$. 

The special cases $x=0$ and $x=1$ correspond to the
standard Hubbard model, albeit with a shift $\pm\Delta$ in the chemical potential. In the case of
constant chemical potential if this shift happens to place the Fermi level below the bottom of the
lower Hubbard band, or above the top of the upper Hubbard band, then we have a "BI", in the sense
that we are dealing with an empty band or completely filled band. 
If the shifted Fermi level crosses the upper or lower Hubbard band, then we are dealing
with a (strongly correlated) metallic state. In this limit, since the lattice is dominated with only one atom,
there is no phase space for randomness, and therefore the AI phases are absent. Once the composition $x$
deviates from these two limits, randomness starts to generate AI phases as well.

\section{Concluding remarks \label{conclusion}}
We have explored the physical effects caused by the simultaneous presence of on-site Coulomb interaction and disorder which is distributed in a 
bimodal form (alloy disorder) on honeycomb lattice within strong-coupling perturbation expansion. In this approach, the inter-site hopping $t$ is considered as the perturbation parameter.  Moreover, the expansion in powers of the hopping $t$ is expressed in terms of local connected correlation functions.
In this paper, we have carried out the perturbative expansion of the auxiliary fermions around the atomic limit up to second order 
in terms of the hopping amplitude. This is already enough
to capture the Mott transition, and the resulting self-energy being local can be integrated into efficient
numerical methods of disordered systems such as transfer matrix method and KPM~\cite{Habibi18}. 
The KPM allows us to efficiently compute the spectral density. The DOS obtained in this way determines whether the system is
gapped or gapless. In a gapless state, the transfer matrix method can be used to calculate the localization length
and its scaling behavior of the quasi particles at the Fermi level.  
Integration of strong-coupling perturbation theory with transfer matrix provides a very powerful and
conclusive tool to determine whether a given state at the Fermi level is an extended (metallic) or (Anderson) localized.

The weakly correlated phases at weak interaction $U$ for fixed chemical potential $\mu=U/2$ is a disordered BI which upon increasing
the correlations becomes AI. When the density is fixed, the life at weak interactions $U$ starts in the AI phase. 
From this point the qualitative behavior of fixed density and fixed chemical potential cases is similar. 
By further increasing the Hubbard $U$ a remarkable metallic phase around $U\approx2\Delta$ emerges which 
can be interpreted as a perfect screening of the disorder by Hubbard interaction. By further increase of the Hubbard $U$, again
an Anderson insulating phase is obtained. Of course in the absence of the alloy energy scale $\Delta$, the Hubbard $U$
on top of a metallic state would stabilize a Mott phase by substantial super-exchange coming from large $U$. 
But when the (disordered) alloy potential is present, it can take advantage of the suppressed charge fluctuations,
and render the metallic phase around $U\approx 2\Delta$ into AI before ultimately the Hubbard $U$ takes over and the system becomes MI. 
Note that a metallic phase is sandwiched between AI phases has not been obtained by mean field theory and DMFT studies~\cite{Balzer05,Lombardo06}.

For $x=0.5$, the metallic phase sandwiched between the two AI can be understood as follows:
At this composition the ordered limit where A and B atoms belong to two sublattices, our system will become the ionic Hubbard
model. Our earlier studies of the ionic Hubbard model indicates a conducting phase between the band and Mott 
insulating phases~\cite{Ebrahimkhas12,ShahnazHafez,Hafez2009CUT}. Strong-coupling perturbation study sheds a new light on this conducting state: 
At $U=2\Delta$~\cite{Adibi16} the divergence of the self-energy $\Gamma$ of auxiliary fermions is responsible for the
formation of gapless state. 
Randomizing the position of $\pm \Delta$ ionic potentials, broadens the $U=2\Delta$ quantum critical conducting phase
of the ionic Hubbard model~\cite{Adibi16} into a region around $U\approx 2\Delta$ where the self-energy $\Gamma$ diverges.
As a result, disorder $\Delta$ and Hubbard $U$ knockout each other in Eq.~\eqref{green'sfunction} and both loos which leaves us with almost a non-interacting electrons
on the lattice. 
The above picture holds for any composition ratio $x$. The width of the AI phase sandwiched between the metallic and MI phase
becomes zero for $x=0$ and $x=1$ and hence the width near these regions must be some power of $x(1-x)$. 

The AI phase in the weak $U$ is a standard AI phase which results from interference.  
The AI phase to the right (larger $U$) side of the metallic phase however results from
the suppression of charge fluctuations by super-exchange mechanism. The later AI phase is narrow
compared to the former AI phase. 
The generic effect of the alloy disorder $\Delta$ is to increase the critical value required for the Mott phase. 

It is interesting that a metallic state together with three known insulating states, namely BI, AI, and MI
exist in the phase diagram of the same model, albeit with a metallic phase in between the two AI.
This calls for the examination of other physical properties of such phases, and comparison of e.g. the
optical absorption spectra to study details of the strongly correlated dynamics resulting from the
competition between the Hubbard $U$, ionic scale $\Delta$ and the randomness. 

\section{acknowledgements}
S.A.J. thanks Iran Science Elites Federation (ISEF) for financial support. 

\appendix

\section{Auxiliary fermion's self-energies \label{self-energy}}
In this Appendix, we present the self-energies of the auxiliary fermions for zeroth and second order diagram of Fig.~\ref{diagram} at an arbitrary temperature $1/\beta$ and chemical potential $\mu=U/2$. Also, we use the abbreviation $u=U/2$.
The mean occupation of each lattice site for a given spin projection at $\mu=u$ is,
\bea
n_{i}=\frac{e^{\beta(u-\epsilon_i)}+e^{-2\beta\epsilon_i}}{Z_i},
\eea
where
$Z_i$ denotes the partition function. For fixed $\mu=u$ the partition function is given by,
\bea
Z_i=1+2\ e^{\beta(u-\epsilon_i)}+e^{-2\beta\epsilon_i}.
\eea
It should be mentioned that in the absence of  magnetic field, the mean occupation for both spins is identical. 

The matrix elements of the self-energy  in zeroth order $\Gamma^{(0)}$ is given by,
\bea
\Gamma_{ij}^{(0)}(i\omega)=\Big(\frac{1- n_{i}}{i\omega-\epsilon_i+u}+\frac{ n_{i}}{i\omega-\epsilon_i-u}\Big)\ \delta_{ij}\label{self-energy-0},
\eea
where $\delta_{ij}$ is Kronecker delta and $i,j$ label the lattice sites. 

The second order self-energy of the auxiliary fermions is,
\bea
\Gamma_{ij}^{(2)}(i\omega)&=&-\delta_{ij}\ \sum_{l}\ V_{il}\ V_{lj}\ \Xi_{il}(i\omega),
\eea
where $i$ and $l$ are nearest neighbor lattice sites connected by hopping. 
In the following we present the $\Xi_{il}(i\omega)$ which is separately calculated in two different 
cases $\epsilon_i=\epsilon_l$ and $\epsilon_i\neq\epsilon_l$. 

In the $\epsilon_i\neq\epsilon_l$ case, the $\Xi_{il}(i\omega)$ is given by,
\begin{widetext}
\bea
\Xi_{il}(i\omega)&=&\Bigg\{
\frac{4\beta u^2\ n_i\ (1-n_i)\ (1-n_l)}{(i\omega-\epsilon_i)^2-u^2}\Big[\frac{n_F(\epsilon_i-u)\ \delta(i\omega-\epsilon_i+u)}{2u\ (\epsilon_i-\epsilon_l)}-\frac{n_F(\epsilon_i+u)\ \delta(i\omega-\epsilon_i-u)}{2u\ (\epsilon_i-\epsilon_l+2u)}\nn\\
&&-\frac{n_F(\epsilon_l-u)\ \delta(i\omega-\epsilon_l+u)}{(\epsilon_i-\epsilon_l)\ (\epsilon_i-\epsilon_l+2u)}  \Big]\nn\\
&&+\frac{4\beta u^2\ n_i\ (1-n_i)\ n_l}{(i\omega-\epsilon_i)^2-u^2}\Big[\frac{n_F(\epsilon_i-u)\ \delta(i\omega-\epsilon_i+u)}{2u\ (\epsilon_i-\epsilon_l-2u)}-\frac{n_F(\epsilon_i+u)\ \delta(i\omega-\epsilon_i-u)}{2u\ (\epsilon_i-\epsilon_l)}-\frac{n_F(\epsilon_l+u)\ \delta(i\omega-\epsilon_l-u)}{(\epsilon_i-\epsilon_l)\ (\epsilon_i-\epsilon_l-2u)} \Big]\nn\\
&&-\frac{\beta\ e^{\beta(u-\epsilon_i)}\ (1-n_l)}{Z_i\ (i\omega-\epsilon_i+u)^2\ (\epsilon_i-\epsilon_l+2u)^2}\Big[(i\omega-\epsilon_l+3u)^2\ n_F(\epsilon_l-u)\ \delta(i\omega-\epsilon_l+u)\nn\\
&&-(i\omega-\epsilon_i+u)^2\ (\epsilon_i-\epsilon_l+2u)\ n_F(\epsilon_i+u)\ \delta'(i\omega-\epsilon_i-u)\nn\\
&&+(i\omega-\epsilon_i+u)\ \delta(i\omega-\epsilon_i-u)\ \Big((i\omega-\epsilon_i+u)\ (\epsilon_i-\epsilon_l+2u)\ n_F'(\epsilon_i+u)-(i\omega+\epsilon_i-2\epsilon_l+5u)\ n_F(\epsilon_i+u)\Big)\Big]\nn\\
&&-\frac{\beta\ e^{\beta(u-\epsilon_i)}\ n_l}{Z_i\ (i\omega-\epsilon_i+u)^2\ (\epsilon_i-\epsilon_l)^2}\Big[(i\omega-\epsilon_l+u)^2\ n_F(\epsilon_l+u)\ \delta(i\omega-\epsilon_l-u)\nn\\
&&-(i\omega-\epsilon_i+u)^2\ (\epsilon_i-\epsilon_l)\ n_F(\epsilon_i+u)\ \delta'(i\omega-\epsilon_i-u)\nn\\
&&+(i\omega-\epsilon_i+u)\ \delta(i\omega-\epsilon_i-u)\ \Big((i\omega-\epsilon_i+u)\ (\epsilon_i-\epsilon_l)\ n_F'(\epsilon_i+u)-(i\omega+\epsilon_i-2\epsilon_l+u)\ n_F(\epsilon_i+u)\Big)\Big]\nn\\
&&-\frac{2\  n_F(\epsilon_i-u)}{(i\omega-\epsilon_i)^2-u^2}\Big(\frac{\epsilon_i-\epsilon_l+2u(n_l-1)}{(\epsilon_i-\epsilon_l)(\epsilon_i-\epsilon_l-2u)}\Big) \Big[\beta u\ n_i\ (1-n_i)+(1-n_i)+\frac{\beta u(e^{-2\beta\epsilon_i}-e^{2\beta(u-\epsilon_i)})}{Z_i^2} \Big] \nn\\
&&+\frac{2\ n_F(\epsilon_i+u)}{(i\omega-\epsilon_i)^2-u^2}\Big(\frac{\epsilon_i-\epsilon_l+2u n_l}{(\epsilon_i-\epsilon_l)(\epsilon_i-\epsilon_l+2u)}\Big)\Big[\beta u\ n_i\ (1-n_i)+n_i+\frac{\beta u(e^{-2\beta\epsilon_i}-e^{2\beta(u-\epsilon_i)})}{Z_i^2} \Big]\nn\\
&&+\frac{4u [\beta\ u\ n_i (1-n_i)+(1-n_i)+\beta u (e^{-2\beta\epsilon_i}-e^{2\beta(u-\epsilon_i)})/Z_i^2 ]}{(\epsilon_i-\epsilon_l)((i\omega-\epsilon_i)^2-u^2)}\Big(\frac{(1-n_l)\ n_F(\epsilon_l-u)}{\epsilon_i-\epsilon_l+2u}+\frac{n_l\ n_F(\epsilon_l+u)}{\epsilon_i-\epsilon_l-2u} \Big)\nn\\
&&-\frac{4 u^2 (2 n_i-1)\ n_F(-i\omega+2\epsilon_i)\ [ i\omega-2\epsilon_i+\epsilon_l+u(1-2n_l)]}{((i\omega-\epsilon_i)^2-u^2)^2\ ((i\omega-2\epsilon_i+\epsilon_l)^2-u^2)}\nn\\
&&+\frac{(2 n_i-1)\ (1-n_l)\ n_F(\epsilon_l-u)}{(i\omega-2\epsilon_i+\epsilon_l-u)}\ \Big(\frac{1}{i\omega-\epsilon_i-u}+\frac{1}{\epsilon_l+\epsilon_i-2u} \Big)^2\nn\\
&&+\frac{(2 n_i-1)\ n_l\ n_F(\epsilon_l+u)}{(i\omega-2\epsilon_i+\epsilon_l+u)}\ \Big(\frac{1}{i\omega-\epsilon_i-u}+\frac{1}{\epsilon_i+\epsilon_l} \Big)^2\nn\\
&&-\frac{n_i\ n_F(\epsilon_i+u) }{(i\omega-\epsilon_i+u)^2}\ \Big(\frac{(1-n_l)(i\omega-\epsilon_l+3u)}{(\epsilon_i-\epsilon_l+2u)^2}+\frac{n_l\  (i\omega-\epsilon_l+u)}{(\epsilon_i-\epsilon_l)^2} \Big)\nn\\
&&+\frac{n_i\ n_F(\epsilon_i+u) }{(i\omega-\epsilon_i-u)^2}\ \Big(\frac{(1-n_l)(i\omega-2\epsilon_i+\epsilon_l-3u)}{(\epsilon_i-\epsilon_l+2u)^2}+\frac{n_l\  (i\omega-2\epsilon_i+\epsilon_l-u)}{(\epsilon_i-\epsilon_l)^2} \Big)\nn\\
&&-\frac{2\ u}{(\epsilon_i-\epsilon_l)\ ((i\omega-\epsilon_i)^2-u^2)}\Big(\frac{n_i (\epsilon_i-\epsilon_l+2u\ n_l)\ n_F'(\epsilon_i+u)}{(\epsilon_i-\epsilon_l+2u)}+\frac{(1-n_i)(\epsilon_i-\epsilon_l+2u\ (n_l-1))\ n_F'(\epsilon_i-u)}{(\epsilon_i-\epsilon_l-2u)} \Big)\nn\\
&&+\frac{(n_i-1)\ n_F(\epsilon_i-u)}{(i\omega-\epsilon_i+u)^2}\Big(\frac{n_l\ (i\omega-2\epsilon_i+\epsilon_l+3u)}{(\epsilon_i-\epsilon_l-2u)^2} +\frac{(1-n_l)\ (i\omega-2\epsilon_i+\epsilon_l+u)}{(\epsilon_i-\epsilon_l)^2} \Big)\nn\\
&&+\frac{(1-n_i)\ n_F(\epsilon_i-u)}{(i\omega-\epsilon_i-u)^2}\Big(\frac{n_l\ (i\omega-\epsilon_l-3u)}{(\epsilon_i-\epsilon_l-2u)^2} +\frac{(1-n_l)\ (i\omega-\epsilon_l-u)}{(\epsilon_i-\epsilon_l)^2} \Big)\nn\\
&&+\frac{(1-n_i)\ (1-n_l)\ n_F(\epsilon_l-u)}{(i\omega-\epsilon_i+u)^2}\Big(\frac{i\omega-\epsilon_l+3u}{(\epsilon_i-\epsilon_l+2u)^2}+\frac{i\omega+\epsilon_l+u}{(\epsilon_i-\epsilon_l)^2}\Big)\nn\\
&&+\frac{(n_l-1)\ n_F(\epsilon_l-u)}{(i\omega-\epsilon_i-u)^2}\Big(\frac{n_i\ (i\omega-2\epsilon_i+\epsilon_l-3u)}{(\epsilon_i-\epsilon_l+2u)^2} +\frac{(1-n_i)\ (i\omega-\epsilon_l-u)}{(\epsilon_i-\epsilon_l)^2} \Big)\nn\\
&&-\frac{n_l\ n_F(\epsilon_l+u)}{(i\omega-\epsilon_i-u)^2}\Big(\frac{(1-n_i)(i\omega-\epsilon_l-3u)}{(\epsilon_i-\epsilon_l-2u)^2}+\frac{n_i\ (i\omega-2\epsilon_i+\epsilon_l-u)}{(\epsilon_i-\epsilon_l)^2} \Big)\nn\\
&&+\frac{(1-n_i)\ n_l\ n_F(\epsilon_l+u)}{(i\omega-\epsilon_i+u)^2}\Big(\frac{i\omega-\epsilon_l+u}{(\epsilon_i-\epsilon_l)^2}+\frac{i\omega-2\epsilon_i+\epsilon_l+3u}{(\epsilon_i-\epsilon_l-2u)^2} \Big)
\Bigg\},
\eea

For $\epsilon_i = \epsilon_l$, $\Xi_{il}$ it reduces to,
\bea
\Xi_{il}(i\omega)&=&\delta_{il}\  \Bigg\{\frac{\beta\ n_i\ (1-n_i)}{(i\omega-\epsilon_i)^2-u^2}\Big[ (1-2 n_i)\ \Big(n_F(\epsilon_i+u)-n_F(\epsilon_i-u)\Big) +2u\ \Big(n_i\ n'_F(\epsilon_i+u)-(1-n_i)\ n'_F(\epsilon_i-u)\Big)\nn\\
&+&(1-2n_i)\Big(\delta(i\omega-\epsilon_i+u)\ n_F(\epsilon_i-u)-\delta(i\omega-\epsilon_i-u)\ n_F(\epsilon_i+u)\Big)\nn\\
&+&2u\ n_i\ \Big(\delta'(i\omega-\epsilon_i-u)\ n_F(\epsilon_i+u)-\delta(i\omega-\epsilon_i-u)\ n'_F(\epsilon_i+u)\Big)\nn\\
&+&2u\ (1-n_i)\ \Big(\delta(i\omega-\epsilon_i+u)\ n'_F(\epsilon_i-u)-\delta'(i\omega-\epsilon_i+u)\ n_F(\epsilon_i-u)\Big)\Big]\nn\\
&+&(2n_i-1)\ (1-n_i)\ \Big[ \frac{n_F(\epsilon_i-u)-n_F(2\epsilon_i-i\omega)}{(i\omega-\epsilon_i-u)^3}+\frac{(i\omega-\epsilon_i+3u)\ n_F(2\epsilon_i-i\omega)}{((i\omega-\epsilon_i)^2-u^2)^2}+\frac{n_F(\epsilon_i+u)}{u\ ((i\omega-\epsilon_i)^2-u^2)}\nn\\
&-&\frac{n_F(\epsilon_i-u)}{u\ (i\omega-\epsilon_i-u)^2}+\frac{n_F(\epsilon_i-u)}{4u^2\ (i\omega-\epsilon_i-u)}+\frac{n'_F(\epsilon_i+u)}{2u\ (i\omega-\epsilon_i+u)}-\frac{(i\omega-\epsilon_i+3u)\ n_F(\epsilon_i+u)}{4u^2\ (i\omega-\epsilon_i+u)^2}\nn\\
&+&\frac{n_F(\epsilon_i+u)-n_F(\epsilon_i-u)}{2u\ (i\omega-\epsilon_i+u)^2}\Big]\nn\\
&+&(2n_i-1)\ n_i\ \frac{n_F(\epsilon_i+u)-n_F(2\epsilon_i-i\omega)}{(i\omega-\epsilon_i+u)}\Big(\frac{1}{(i\omega-\epsilon_i+u)^2}+\frac{1}{(i\omega-\epsilon_i-u)^2}-\frac{2}{((i\omega-\epsilon_i)^2-u^2)}\Big)\nn\\
&+&(2n_i-1)\ n_i\ \Big[\frac{2\ n'_F(\epsilon_i+u)}{((i\omega-\epsilon_i)^2-u^2)}-\frac{n'_F(\epsilon_i+u)}{(i\omega-\epsilon_i+u)^2}+\frac{n''_F(\epsilon_i+u)}{2(i\omega-\epsilon_i+u)}\Big]\nn\\
&+&\frac{\beta\ (e^{-2\beta\epsilon_i}-e^{2\beta(u-\epsilon_i)})}{Z_i^2\ ((i\omega-\epsilon_i)^2-u^2)}\ \Big[ (2n_i-1)\ \Big(n_F(\epsilon_i-u)-n_F(\epsilon_i+u)\Big)+2u\ \Big(n_i\ n'_F(\epsilon_i+u)-(1-n_i)\ n'_F(\epsilon_i-u)\Big)\Big]\nn\\
&+&\frac{(1-n_i)}{u\ ((i\omega-\epsilon_i)^2-u^2)}\ \Big[(2n_i-1)\ (n_F(\epsilon_i-u)-n_F(\epsilon_i+u))+2n_i\ n'_F(\epsilon_i+u)-2(1-n_i)\ n'_F(\epsilon_i-u)\Big]\nn\\
&+&\frac{n_i\ (n_i-1)\ n'_F(\epsilon_i+u)+(n_i-1)^2\ n'_F(\epsilon_i-u)}{(i\omega-\epsilon_i+u)^2}+\frac{-n_i^2\ n'_F(\epsilon_i+u)+(n_i-1)^2\ n'_F(\epsilon_i-u)}{(i\omega-\epsilon_i-u)^2}\nn\\
&+&\frac{n_i\ (1-n_i)\ n''_F(\epsilon_i+u)+(n_i-1)^2\ n''_F(\epsilon_i-u)}{(i\omega-\epsilon_i+u)}-\frac{n_i^2\ n''_F(\epsilon_i+u)+(n_i-1)^2\ n''_F(\epsilon_i-u)}{(i\omega-\epsilon_i-u)}\nn\\
&+&\frac{(1-n_i)}{4u^2}\ \Big(n_F(\epsilon_i-u)-n_F(\epsilon_i+u)+2u\ n'_F(\epsilon_i+u)\Big)\ \Big(\frac{1-n_i}{i\omega-\epsilon_i+u}-\frac{n_i}{i\omega-\epsilon_i-u}\Big)\nn\\
&+&\frac{n_i\ (1-n_i)}{4u^2}\ \Big(n_F(\epsilon_i+u)-n_F(\epsilon_i-u)-2u\ n'_F(\epsilon_i-u)\Big)\ \Big(\frac{1}{i\omega-\epsilon_i+u}-\frac{1}{i\omega-\epsilon_i-u}\Big)\nn\\
&-&\frac{\beta\ e^{\beta(u-\epsilon_i)}\ (1-n_i)}{4u\ Z_i\ (i\omega-\epsilon_i+u)^2}\Big[(i\omega-\epsilon_i+3u)^2\  n_F(\epsilon_i-u)\ \delta(i\omega-\epsilon_i+u)\nn\\
&-&(i\omega-\epsilon_i+u)\ (i\omega-\epsilon_i+5u)\ \ n_F(\epsilon_i+u)\ \delta(i\omega-\epsilon_i-u)\nn\\
&+&2u\ (i\omega-\epsilon_i+u)^2\  \Big(n'_F(\epsilon_i+u)\ \delta'(i\omega-\epsilon_i-u)-n_F(\epsilon_i+u)\ \delta'(i\omega-\epsilon_i-u)\Big)\Big]\nn\\
&-&\frac{\beta\ e^{\beta(u-\epsilon_i)}\ n_i}{2\ Z_i\ (i\omega-\epsilon_i+u)^2}\Big[2n_F(\epsilon_i+u)\ \delta(i\omega-\epsilon_i-u)-2(i\omega-\epsilon_i+u)^2\ \delta'(i\omega-\epsilon_i-u)\ n'_F(\epsilon_i+u)\nn\\
&+&n_F(\epsilon_i+u)\ (i\omega-\epsilon_i+u)\ \Big(4\delta'(i\omega-\epsilon_i-u)+(i\omega-\epsilon_i+u)\ \delta''(i\omega-\epsilon_i-u)\Big)\nn\\
&+&(i\omega-\epsilon_i+u)\ \delta(i\omega-\epsilon_i-u)\Big((i\omega-\epsilon_i+u)\ n''_F(\epsilon_i+u)-4n'_F(\epsilon_i+u)\Big)\Big]\Bigg\},
\eea
\end{widetext}
where $n_F(x)=\frac{1}{\exp(\beta x)+1}$ and $\delta(x)$ denote the Fermi-Dirac distribution and delta function, respectively. 
Further, we represent the first derivative of these functions with $n'_F(x)$ and $\delta'(x)$ and the second derivatives with $n''_F(x)$ and $\delta''(x)$. 

 \section{Kernel polynomial method}
 \label{kpm}
 The KPM is a stochastic approach which is based on the expansion of any spectral function into a finite series of Chebyshev (or any
other complete set of orthonormal) polynomials \cite{KPM,Habibi13,Habibi2018Resilience}. The expansion coefficients are computed through an efficient recursion relation which involves sparse matrix and vector multiplications with Hamiltonian $H$, followed by possible regularization~\cite{SotaRKPM,Amini09}. 
On the other hand, the arguments of  Chebyshev polynomials does not exceed $1$.
In consequence, expanding the Hamiltonian $H$,  which its  eigenvalues $E$ are in range $[E_{min},E_{max}]$, in Chebyshev polynomials requires to rescale to $\hat{H}(\varepsilon)$ where  $\varepsilon\in[-1,1]$. Also, $\hat{H}(\varepsilon)$ and  $\varepsilon$ are defined as $\hat{H}=(H-b)/a$ and $\varepsilon=(E-b)/a$  where $b=(E_{max}+E_{min})/2 $ and  $a=(E_{max}-E_{min})/2$.

The DOS can be expanded as follows,
    \begin{eqnarray} 
    \hat{\rho}(\varepsilon)=\frac{1}{\pi \sqrt{1-\varepsilon^2}}(\mu_0\ g_0+2\sum_{m=1}^{N_c}\mu_m\  g_m\  T_m(\varepsilon) ),
    \label{dos-kpm} 
    \end{eqnarray} 
where $T_m(\varepsilon)=\cos(m \arccos(\varepsilon))$  are the $m$-th Chebyshev polynomials,  $g_m$  are attenuation factors which minimize
the Gibbs oscillations arising from terminating the expansion  in a  finite order. $N_c$ denotes the cut off on the expansion order. $\mu_m$  are Chebyshev moments which expressed as,
\bea
\mu_m=\frac{1}{N_s} \sum_{r=1}^{N_s} \langle\phi_r\vert T_m(\hat{H}) \vert \phi_r \rangle,
\eea
where $\phi_r$  are random single-particle states and $N_s$ is the number of random states used in numerical calculations. To calculate matrix elements of $T_m(\hat{H})$ we use the recurrence relation of Chebyshev polynomials, namely,   $T_m (\hat{H})= 2\hat{H} T_{m-1}(\hat{H}) - T_{m-2}(\hat{H})$ with  initial conditions $T_{1}(\hat{H})=\hat{H}$ and $T_{0}(\hat{H})=1$.

To apply the above procedure for calculating DOS to Green's function \eqref{green'sfunction}, we utilize the following equation,
\begin{align}
    \rho'(\omega)=-\frac{1}{\pi}\lim_{\eta\to 0}\mathrm{Im}\frac{1}{E+i\eta+\Gamma^{-1}(i\omega)-V}\Big|_{E=0}.
\end{align}
So the DOS in Eq.~\eqref{dos-kpm} can be rewritten as,
\begin{align}
        \hat{\rho}'(\omega')=\frac{1}{\pi \sqrt{1-\epsilon^2}}\ \Big(\mu_0\  g_0+2\sum_{m=1}^{N_c}\mu_m(\omega')\  g_m\  T_m(\epsilon) \Big)\Big|_{\epsilon=0},
\end{align}
where $\mu_m(\omega')$ are the generalized  Chebyshev moments in which $H$ is considered as $\Gamma^{-1}(\omega)-V$. Furthermore, $\omega'$ and $\hat{H}$ are rescaled  $\omega$ and $H$, respectively. To compute $\mu_m(\omega')$, we need to calculate $\mu_m$ for every $\omega'$ which is computationally expensive. 
Therefore, MPICH is employed to paralleliz our program. Additionally, owing to divergences of  $\Gamma^{-1}(\omega)$ for some values in disorder distribution 
and the spectral broadening, we need to choose large cutoff $N_c=15000$, $N_s=5$, and calculate its average on $100$ configurations of disorder 
to obtain well converged values of DOS $\rho'(\omega')$ at $E=0$.

  \begin{figure}[b!]
    \centering
    \includegraphics[width=0.9\linewidth]{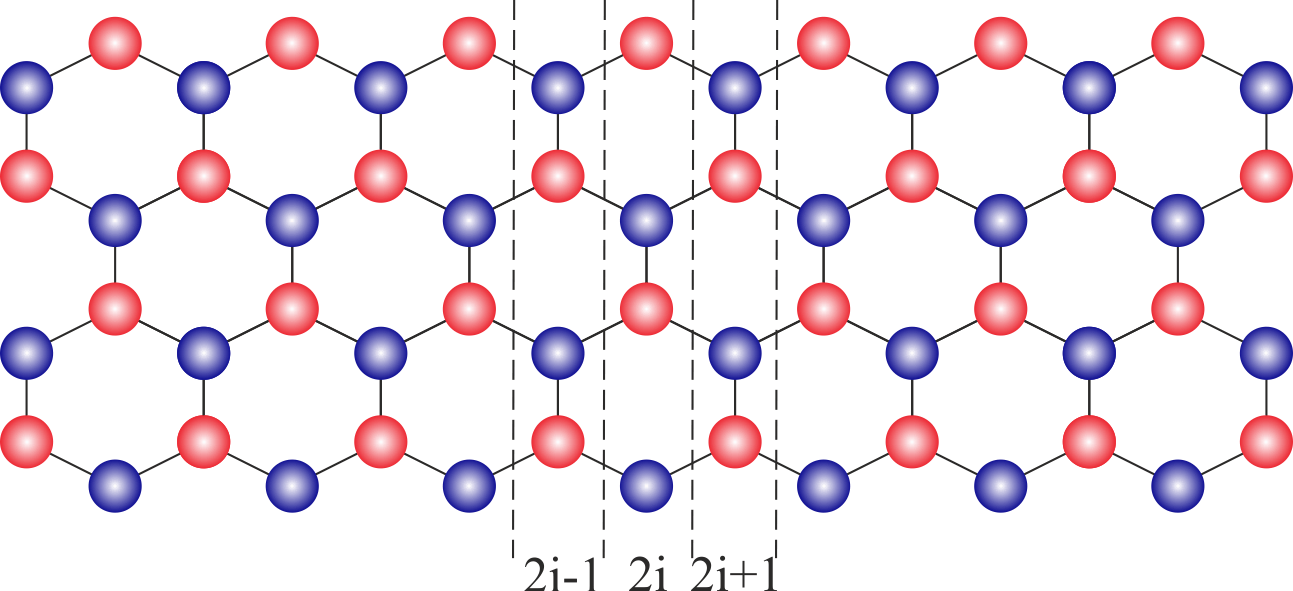}
    \caption{Honeycomb lattice with zigzag edge as the transport direction and width $M=4$ and length $L=15$. }
    \label{fig:TM}
\end{figure}

 \section{Transfer matrix method\label{TMM}}
In this section, we describe the transfer matrix method employed for computing the localization length. 
In this method we use quasi-one dimensional Schr\"{o}dinger equation $H\vec{\Psi }_i=E\vec{\Psi }_i$ which can be rewritten as,
\begin{align}
    \mathbf{V_{i,i-1}^*}\vec{\Psi }_{i-1}+\mathbf{H_{i}}\vec{\Psi }_i+\mathbf{V_{i,i+1}}\vec{\Psi }_{i+1}=E \vec{\Psi }_i,
\end{align}
where $\mathbf{V_{i,i+1}}$ denotes hopping matrix from block $i$ to $i+1$ and $\mathbf{H_{i}}$ is the on-site matrix for block $i$.
The above wave equation can be expressed by the following matrix,
\begin{align}
    \left(
    \begin{array}{c}
        \vec{\Psi }_{i+1} \\
        \vec{\Psi }_{i} 
    \end{array}
    \right) =
    \mathbf{T_{i+1,i}}
    \left(
    \begin{array}{c}
        \vec{\Psi }_{i} \\
        \vec{\Psi }_{i-1} 
    \end{array}\right),\label{TMM-eq1}
\end{align}
where
\begin{align}
    \mathbf{T_{i+1,i}}=\left(
    \begin{array}{cc}
        \mathbf{V^{-1}_{i,i+1}} \left(E \mathbf{I}-\mathbf{H_{i}}\right) &- \mathbf{V^{-1}_{i,i+1} V_{i,i-1}^*}\\
        \\
        \mathbf{I} & \mathbf{0} \\
    \end{array}
    \right).
\end{align}
Indeed, matrix equation~\eqref{TMM-eq1} provides us a recursive procedure to calculate the wave function $\vec{\Psi }_i$ of the $i$-th slice along the transfer direction. In this paper, we used zigzag Graphene with periodic boundary condition as depicted in Fig.~\ref{fig:TM}. Assuming $M$ as the width of the system, vector elements are $M$-by-$M$ matrices whereas $\mathbf{T}$ is $2M$-by-$2M$ matrix.

Oseledecs theorem \cite{Oseledec} states that by defining the product of the transfer matrices as $\mathbf{\Gamma_N=T_{N+1,N}\ T_{N,N-1}\cdots T_{2,1}}$  the eigenvalues of $(\mathbf{\Gamma_N^{\dagger}\ \Gamma_N})^{1/{2N}}$ in thermodynamic limit converge to fixed values $e^{\pm\gamma_m}$ where $\gamma_m$ with $m=1,\cdots, M$ are Lyapunov exponents.
The localization length $\lambda$ is computed by  minimum Lyapunov exponent as the largest decaying length
\begin{align}
    \lambda=\frac{1}{\gamma_{min}}.
\end{align}
The numerical details to compute the smallest positive Lyapunov exponent is presented in Ref. \cite{MacKinnon81,MacKinnon83,habibi2018topological}.
In employing the transfer matrix method, $N$ is chosen in such a way that localization length converges. 

\bibliographystyle{apsrev4-1}
\bibliography{Refs}

\begin{thebibliography}{50}%
\makeatletter
\providecommand \@ifxundefined [1]{%
 \@ifx{#1\undefined}
}%
\providecommand \@ifnum [1]{%
 \ifnum #1\expandafter \@firstoftwo
 \else \expandafter \@secondoftwo
 \fi
}%
\providecommand \@ifx [1]{%
 \ifx #1\expandafter \@firstoftwo
 \else \expandafter \@secondoftwo
 \fi
}%
\providecommand \natexlab [1]{#1}%
\providecommand \enquote  [1]{``#1''}%
\providecommand \bibnamefont  [1]{#1}%
\providecommand \bibfnamefont [1]{#1}%
\providecommand \citenamefont [1]{#1}%
\providecommand \href@noop [0]{\@secondoftwo}%
\providecommand \href [0]{\begingroup \@sanitize@url \@href}%
\providecommand \@href[1]{\@@startlink{#1}\@@href}%
\providecommand \@@href[1]{\endgroup#1\@@endlink}%
\providecommand \@sanitize@url [0]{\catcode `\\12\catcode `\$12\catcode
  `\&12\catcode `\#12\catcode `\^12\catcode `\_12\catcode `\%12\relax}%
\providecommand \@@startlink[1]{}%
\providecommand \@@endlink[0]{}%
\providecommand \url  [0]{\begingroup\@sanitize@url \@url }%
\providecommand \@url [1]{\endgroup\@href {#1}{\urlprefix }}%
\providecommand \urlprefix  [0]{URL }%
\providecommand \Eprint [0]{\href }%
\providecommand \doibase [0]{http://dx.doi.org/}%
\providecommand \selectlanguage [0]{\@gobble}%
\providecommand \bibinfo  [0]{\@secondoftwo}%
\providecommand \bibfield  [0]{\@secondoftwo}%
\providecommand \translation [1]{[#1]}%
\providecommand \BibitemOpen [0]{}%
\providecommand \bibitemStop [0]{}%
\providecommand \bibitemNoStop [0]{.\EOS\space}%
\providecommand \EOS [0]{\spacefactor3000\relax}%
\providecommand \BibitemShut  [1]{\csname bibitem#1\endcsname}%
\let\auto@bib@innerbib\@empty
\bibitem [{\citenamefont {Sorella}\ and\ \citenamefont
  {Tosatti}(1992)}]{sorella1992}%
  \BibitemOpen
  \bibfield  {author} {\bibinfo {author} {\bibfnamefont {S.}~\bibnamefont
  {Sorella}}\ and\ \bibinfo {author} {\bibfnamefont {E.}~\bibnamefont
  {Tosatti}},\ }\href {\doibase 10.1209/0295-5075/19/8/007} {\bibfield
  {journal} {\bibinfo  {journal} {Eur. Phys. Lett.}\ }\textbf {\bibinfo
  {volume} {19}},\ \bibinfo {pages} {699} (\bibinfo {year} {1992})}\BibitemShut
  {NoStop}%
\bibitem [{\citenamefont {Jafari}(2009)}]{jafari2009}%
  \BibitemOpen
  \bibfield  {author} {\bibinfo {author} {\bibfnamefont {S.~A.}\ \bibnamefont
  {Jafari}},\ }\href {\doibase 10.1140/epjb/e2009-00128-1} {\bibfield
  {journal} {\bibinfo  {journal} {Eur. Phys. J. B}\ }\textbf {\bibinfo {volume}
  {68}},\ \bibinfo {pages} {537} (\bibinfo {year} {2009})}\BibitemShut
  {NoStop}%
\bibitem [{\citenamefont {Wu}\ \emph {et~al.}(2010)\citenamefont {Wu},
  \citenamefont {Chen}, \citenamefont {Tao}, \citenamefont {Tong},\ and\
  \citenamefont {Liu}}]{Wu10}%
  \BibitemOpen
  \bibfield  {author} {\bibinfo {author} {\bibfnamefont {W.}~\bibnamefont
  {Wu}}, \bibinfo {author} {\bibfnamefont {Y.-H.}\ \bibnamefont {Chen}},
  \bibinfo {author} {\bibfnamefont {H.-S.}\ \bibnamefont {Tao}}, \bibinfo
  {author} {\bibfnamefont {N.-H.}\ \bibnamefont {Tong}}, \ and\ \bibinfo
  {author} {\bibfnamefont {W.-M.}\ \bibnamefont {Liu}},\ }\href {\doibase
  10.1103/PhysRevB.82.245102} {\bibfield  {journal} {\bibinfo  {journal} {Phys.
  Rev. B}\ }\textbf {\bibinfo {volume} {82}},\ \bibinfo {pages} {245102}
  (\bibinfo {year} {2010})}\BibitemShut {NoStop}%
\bibitem [{\citenamefont {He}\ and\ \citenamefont {Lu}(2012)}]{He12}%
  \BibitemOpen
  \bibfield  {author} {\bibinfo {author} {\bibfnamefont {R.-Q.}\ \bibnamefont
  {He}}\ and\ \bibinfo {author} {\bibfnamefont {Z.-Y.}\ \bibnamefont {Lu}},\
  }\href {\doibase 10.1103/PhysRevB.86.045105} {\bibfield  {journal} {\bibinfo
  {journal} {Phys. Rev. B}\ }\textbf {\bibinfo {volume} {86}},\ \bibinfo
  {pages} {045105} (\bibinfo {year} {2012})}\BibitemShut {NoStop}%
\bibitem [{\citenamefont {Hassan}\ and\ \citenamefont
  {S\'en\'echal}(2013)}]{Hassan13}%
  \BibitemOpen
  \bibfield  {author} {\bibinfo {author} {\bibfnamefont {S.~R.}\ \bibnamefont
  {Hassan}}\ and\ \bibinfo {author} {\bibfnamefont {D.}~\bibnamefont
  {S\'en\'echal}},\ }\href {\doibase 10.1103/PhysRevLett.110.096402} {\bibfield
   {journal} {\bibinfo  {journal} {Phys. Rev. Lett.}\ }\textbf {\bibinfo
  {volume} {110}},\ \bibinfo {pages} {096402} (\bibinfo {year}
  {2013})}\BibitemShut {NoStop}%
\bibitem [{\citenamefont {Otsuka}\ \emph {et~al.}(2016)\citenamefont {Otsuka},
  \citenamefont {Yunoki},\ and\ \citenamefont {Sorella}}]{Otsuka16}%
  \BibitemOpen
  \bibfield  {author} {\bibinfo {author} {\bibfnamefont {Y.}~\bibnamefont
  {Otsuka}}, \bibinfo {author} {\bibfnamefont {S.}~\bibnamefont {Yunoki}}, \
  and\ \bibinfo {author} {\bibfnamefont {S.}~\bibnamefont {Sorella}},\ }\href
  {\doibase 10.1103/PhysRevX.6.011029} {\bibfield  {journal} {\bibinfo
  {journal} {Phys. Rev. X}\ }\textbf {\bibinfo {volume} {6}},\ \bibinfo {pages}
  {011029} (\bibinfo {year} {2016})}\BibitemShut {NoStop}%
\bibitem [{\citenamefont {Adibi}\ and\ \citenamefont {Jafari}(2016)}]{Adibi16}%
  \BibitemOpen
  \bibfield  {author} {\bibinfo {author} {\bibfnamefont {E.}~\bibnamefont
  {Adibi}}\ and\ \bibinfo {author} {\bibfnamefont {S.~A.}\ \bibnamefont
  {Jafari}},\ }\href {\doibase 10.1103/PhysRevB.93.075122} {\bibfield
  {journal} {\bibinfo  {journal} {Phys. Rev. B}\ }\textbf {\bibinfo {volume}
  {93}},\ \bibinfo {pages} {075122} (\bibinfo {year} {2016})}\BibitemShut
  {NoStop}%
\bibitem [{\citenamefont {Mott}(1990)}]{MottBook}%
  \BibitemOpen
  \bibfield  {author} {\bibinfo {author} {\bibfnamefont {N.}~\bibnamefont
  {Mott}},\ }\href {\doibase 10.1201/b12795} {\emph {\bibinfo {title}
  {Metal-Insulator Transitions}}}\ (\bibinfo  {publisher} {{CRC} Press},\
  \bibinfo {year} {1990})\BibitemShut {NoStop}%
\bibitem [{\citenamefont {Garg}\ \emph {et~al.}(2006)\citenamefont {Garg},
  \citenamefont {Krishnamurthy},\ and\ \citenamefont {Randeria}}]{Garg06}%
  \BibitemOpen
  \bibfield  {author} {\bibinfo {author} {\bibfnamefont {A.}~\bibnamefont
  {Garg}}, \bibinfo {author} {\bibfnamefont {H.~R.}\ \bibnamefont
  {Krishnamurthy}}, \ and\ \bibinfo {author} {\bibfnamefont {M.}~\bibnamefont
  {Randeria}},\ }\href {\doibase 10.1103/PhysRevLett.97.046403} {\bibfield
  {journal} {\bibinfo  {journal} {Phys. Rev. Lett.}\ }\textbf {\bibinfo
  {volume} {97}},\ \bibinfo {pages} {046403} (\bibinfo {year}
  {2006})}\BibitemShut {NoStop}%
\bibitem [{\citenamefont {Paris}\ \emph
  {et~al.}(2007{\natexlab{a}})\citenamefont {Paris}, \citenamefont {Bouadim},
  \citenamefont {Hebert}, \citenamefont {Batrouni},\ and\ \citenamefont
  {Scalettar}}]{Paris-IHM07}%
  \BibitemOpen
  \bibfield  {author} {\bibinfo {author} {\bibfnamefont {N.}~\bibnamefont
  {Paris}}, \bibinfo {author} {\bibfnamefont {K.}~\bibnamefont {Bouadim}},
  \bibinfo {author} {\bibfnamefont {F.}~\bibnamefont {Hebert}}, \bibinfo
  {author} {\bibfnamefont {G.~G.}\ \bibnamefont {Batrouni}}, \ and\ \bibinfo
  {author} {\bibfnamefont {R.~T.}\ \bibnamefont {Scalettar}},\ }\href {\doibase
  10.1103/PhysRevLett.98.046403} {\bibfield  {journal} {\bibinfo  {journal}
  {Phys. Rev. Lett.}\ }\textbf {\bibinfo {volume} {98}},\ \bibinfo {pages}
  {046403} (\bibinfo {year} {2007}{\natexlab{a}})}\BibitemShut {NoStop}%
\bibitem [{\citenamefont {Bouadim}\ \emph {et~al.}(2007)\citenamefont
  {Bouadim}, \citenamefont {Paris}, \citenamefont {H\'ebert}, \citenamefont
  {Batrouni},\ and\ \citenamefont {Scalettar}}]{Bouadim07}%
  \BibitemOpen
  \bibfield  {author} {\bibinfo {author} {\bibfnamefont {K.}~\bibnamefont
  {Bouadim}}, \bibinfo {author} {\bibfnamefont {N.}~\bibnamefont {Paris}},
  \bibinfo {author} {\bibfnamefont {F.}~\bibnamefont {H\'ebert}}, \bibinfo
  {author} {\bibfnamefont {G.~G.}\ \bibnamefont {Batrouni}}, \ and\ \bibinfo
  {author} {\bibfnamefont {R.~T.}\ \bibnamefont {Scalettar}},\ }\href {\doibase
  10.1103/PhysRevB.76.085112} {\bibfield  {journal} {\bibinfo  {journal} {Phys.
  Rev. B}\ }\textbf {\bibinfo {volume} {76}},\ \bibinfo {pages} {085112}
  (\bibinfo {year} {2007})}\BibitemShut {NoStop}%
\bibitem [{\citenamefont {Craco}\ \emph {et~al.}(2008)\citenamefont {Craco},
  \citenamefont {Lombardo}, \citenamefont {Hayn}, \citenamefont {Japaridze},\
  and\ \citenamefont {M\"uller-Hartmann}}]{Craco08}%
  \BibitemOpen
  \bibfield  {author} {\bibinfo {author} {\bibfnamefont {L.}~\bibnamefont
  {Craco}}, \bibinfo {author} {\bibfnamefont {P.}~\bibnamefont {Lombardo}},
  \bibinfo {author} {\bibfnamefont {R.}~\bibnamefont {Hayn}}, \bibinfo {author}
  {\bibfnamefont {G.~I.}\ \bibnamefont {Japaridze}}, \ and\ \bibinfo {author}
  {\bibfnamefont {E.}~\bibnamefont {M\"uller-Hartmann}},\ }\href {\doibase
  10.1103/PhysRevB.78.075121} {\bibfield  {journal} {\bibinfo  {journal} {Phys.
  Rev. B}\ }\textbf {\bibinfo {volume} {78}},\ \bibinfo {pages} {075121}
  (\bibinfo {year} {2008})}\BibitemShut {NoStop}%
\bibitem [{\citenamefont {Ebrahimkhas}\ and\ \citenamefont
  {Jafari}(2012)}]{Ebrahimkhas12}%
  \BibitemOpen
  \bibfield  {author} {\bibinfo {author} {\bibfnamefont {M.}~\bibnamefont
  {Ebrahimkhas}}\ and\ \bibinfo {author} {\bibfnamefont {S.~A.}\ \bibnamefont
  {Jafari}},\ }\href {\doibase 10.1209/0295-5075/98/27009} {\bibfield
  {journal} {\bibinfo  {journal} {Eur. Phys. Lett.}\ }\textbf {\bibinfo
  {volume} {98}},\ \bibinfo {pages} {27009} (\bibinfo {year}
  {2012})}\BibitemShut {NoStop}%
\bibitem [{\citenamefont {Denteneer}\ \emph {et~al.}(1999)\citenamefont
  {Denteneer}, \citenamefont {Scalettar},\ and\ \citenamefont
  {Trivedi}}]{Denteneer99}%
  \BibitemOpen
  \bibfield  {author} {\bibinfo {author} {\bibfnamefont {P.~J.~H.}\
  \bibnamefont {Denteneer}}, \bibinfo {author} {\bibfnamefont {R.~T.}\
  \bibnamefont {Scalettar}}, \ and\ \bibinfo {author} {\bibfnamefont
  {N.}~\bibnamefont {Trivedi}},\ }\href {\doibase 10.1103/PhysRevLett.83.4610}
  {\bibfield  {journal} {\bibinfo  {journal} {Phys. Rev. Lett.}\ }\textbf
  {\bibinfo {volume} {83}},\ \bibinfo {pages} {4610} (\bibinfo {year}
  {1999})}\BibitemShut {NoStop}%
\bibitem [{\citenamefont {Denteneer}\ and\ \citenamefont
  {Scalettar}(2003)}]{Denteneer03}%
  \BibitemOpen
  \bibfield  {author} {\bibinfo {author} {\bibfnamefont {P.~J.~H.}\
  \bibnamefont {Denteneer}}\ and\ \bibinfo {author} {\bibfnamefont {R.~T.}\
  \bibnamefont {Scalettar}},\ }\href {\doibase 10.1103/PhysRevLett.90.246401}
  {\bibfield  {journal} {\bibinfo  {journal} {Phys. Rev. Lett.}\ }\textbf
  {\bibinfo {volume} {90}},\ \bibinfo {pages} {246401} (\bibinfo {year}
  {2003})}\BibitemShut {NoStop}%
\bibitem [{\citenamefont {Chakraborty}\ \emph {et~al.}(2007)\citenamefont
  {Chakraborty}, \citenamefont {Denteneer},\ and\ \citenamefont
  {Scalettar}}]{Chakraborty07}%
  \BibitemOpen
  \bibfield  {author} {\bibinfo {author} {\bibfnamefont {P.~B.}\ \bibnamefont
  {Chakraborty}}, \bibinfo {author} {\bibfnamefont {P.~J.~H.}\ \bibnamefont
  {Denteneer}}, \ and\ \bibinfo {author} {\bibfnamefont {R.~T.}\ \bibnamefont
  {Scalettar}},\ }\href {\doibase 10.1103/PhysRevB.75.125117} {\bibfield
  {journal} {\bibinfo  {journal} {Phys. Rev. B}\ }\textbf {\bibinfo {volume}
  {75}},\ \bibinfo {pages} {125117} (\bibinfo {year} {2007})}\BibitemShut
  {NoStop}%
\bibitem [{\citenamefont {Heidarian}\ and\ \citenamefont
  {Trivedi}(2004)}]{Heidarian04}%
  \BibitemOpen
  \bibfield  {author} {\bibinfo {author} {\bibfnamefont {D.}~\bibnamefont
  {Heidarian}}\ and\ \bibinfo {author} {\bibfnamefont {N.}~\bibnamefont
  {Trivedi}},\ }\href {\doibase 10.1103/PhysRevLett.93.126401} {\bibfield
  {journal} {\bibinfo  {journal} {Phys. Rev. Lett.}\ }\textbf {\bibinfo
  {volume} {93}},\ \bibinfo {pages} {126401} (\bibinfo {year}
  {2004})}\BibitemShut {NoStop}%
\bibitem [{\citenamefont {Henseler}\ \emph {et~al.}(2008)\citenamefont
  {Henseler}, \citenamefont {Kroha},\ and\ \citenamefont
  {Shapiro}}]{Henseler08}%
  \BibitemOpen
  \bibfield  {author} {\bibinfo {author} {\bibfnamefont {P.}~\bibnamefont
  {Henseler}}, \bibinfo {author} {\bibfnamefont {J.}~\bibnamefont {Kroha}}, \
  and\ \bibinfo {author} {\bibfnamefont {B.}~\bibnamefont {Shapiro}},\ }\href
  {\doibase 10.1103/physrevb.78.235116} {\bibfield  {journal} {\bibinfo
  {journal} {Phys. Rev. B}\ }\textbf {\bibinfo {volume} {78}},\ \bibinfo
  {pages} {235116} (\bibinfo {year} {2008})}\BibitemShut {NoStop}%
\bibitem [{\citenamefont {Song}\ \emph {et~al.}(2008)\citenamefont {Song},
  \citenamefont {Wortis},\ and\ \citenamefont {Atkinson}}]{Atkinson08}%
  \BibitemOpen
  \bibfield  {author} {\bibinfo {author} {\bibfnamefont {Y.}~\bibnamefont
  {Song}}, \bibinfo {author} {\bibfnamefont {R.}~\bibnamefont {Wortis}}, \ and\
  \bibinfo {author} {\bibfnamefont {W.~A.}\ \bibnamefont {Atkinson}},\ }\href
  {\doibase 10.1103/physrevb.77.054202} {\bibfield  {journal} {\bibinfo
  {journal} {Phys. Rev. B}\ }\textbf {\bibinfo {volume} {77}},\ \bibinfo
  {pages} {054202} (\bibinfo {year} {2008})}\BibitemShut {NoStop}%
\bibitem [{\citenamefont {Habibi}\ \emph
  {et~al.}(2018{\natexlab{a}})\citenamefont {Habibi}, \citenamefont {Adibi},\
  and\ \citenamefont {Jafari}}]{Habibi18}%
  \BibitemOpen
  \bibfield  {author} {\bibinfo {author} {\bibfnamefont {A.}~\bibnamefont
  {Habibi}}, \bibinfo {author} {\bibfnamefont {E.}~\bibnamefont {Adibi}}, \
  and\ \bibinfo {author} {\bibfnamefont {S.~A.}\ \bibnamefont {Jafari}},\
  }\href {https://arxiv.org/abs/1806.07969} {\bibfield  {journal} {\bibinfo
  {journal} {arXiv preprint arXiv:1806.07969}\ } (\bibinfo {year}
  {2018}{\natexlab{a}})}\BibitemShut {NoStop}%
\bibitem [{\citenamefont {Balzer}\ and\ \citenamefont
  {Potthoff}(2005)}]{Balzer05}%
  \BibitemOpen
  \bibfield  {author} {\bibinfo {author} {\bibfnamefont {M.}~\bibnamefont
  {Balzer}}\ and\ \bibinfo {author} {\bibfnamefont {M.}~\bibnamefont
  {Potthoff}},\ }\href {\doibase https://doi.org/10.1016/j.physb.2005.01.221}
  {\bibfield  {journal} {\bibinfo  {journal} {Physica B: Condensed Matter}\
  }\textbf {\bibinfo {volume} {359}},\ \bibinfo {pages} {768} (\bibinfo {year}
  {2005})}\BibitemShut {NoStop}%
\bibitem [{\citenamefont {Lombardo}\ \emph {et~al.}(2006)\citenamefont
  {Lombardo}, \citenamefont {Hayn},\ and\ \citenamefont
  {Japaridze}}]{Lombardo06}%
  \BibitemOpen
  \bibfield  {author} {\bibinfo {author} {\bibfnamefont {P.}~\bibnamefont
  {Lombardo}}, \bibinfo {author} {\bibfnamefont {R.}~\bibnamefont {Hayn}}, \
  and\ \bibinfo {author} {\bibfnamefont {G.~I.}\ \bibnamefont {Japaridze}},\
  }\href {\doibase 10.1103/PhysRevB.74.085116} {\bibfield  {journal} {\bibinfo
  {journal} {Phys. Rev. B}\ }\textbf {\bibinfo {volume} {74}},\ \bibinfo
  {pages} {085116} (\bibinfo {year} {2006})}\BibitemShut {NoStop}%
\bibitem [{\citenamefont {Byczuk}\ \emph {et~al.}(2003)\citenamefont {Byczuk},
  \citenamefont {Ulmke},\ and\ \citenamefont {Vollhardt}}]{Byczuk03}%
  \BibitemOpen
  \bibfield  {author} {\bibinfo {author} {\bibfnamefont {K.}~\bibnamefont
  {Byczuk}}, \bibinfo {author} {\bibfnamefont {M.}~\bibnamefont {Ulmke}}, \
  and\ \bibinfo {author} {\bibfnamefont {D.}~\bibnamefont {Vollhardt}},\ }\href
  {\doibase 10.1103/PhysRevLett.90.196403} {\bibfield  {journal} {\bibinfo
  {journal} {Phys. Rev. Lett.}\ }\textbf {\bibinfo {volume} {90}},\ \bibinfo
  {pages} {196403} (\bibinfo {year} {2003})}\BibitemShut {NoStop}%
\bibitem [{\citenamefont {Byczuk}\ \emph {et~al.}(2004)\citenamefont {Byczuk},
  \citenamefont {Hofstetter},\ and\ \citenamefont {Vollhardt}}]{Byczuk04}%
  \BibitemOpen
  \bibfield  {author} {\bibinfo {author} {\bibfnamefont {K.}~\bibnamefont
  {Byczuk}}, \bibinfo {author} {\bibfnamefont {W.}~\bibnamefont {Hofstetter}},
  \ and\ \bibinfo {author} {\bibfnamefont {D.}~\bibnamefont {Vollhardt}},\
  }\href {\doibase 10.1103/PhysRevB.69.045112} {\bibfield  {journal} {\bibinfo
  {journal} {Phys. Rev. B}\ }\textbf {\bibinfo {volume} {69}},\ \bibinfo
  {pages} {045112} (\bibinfo {year} {2004})}\BibitemShut {NoStop}%
\bibitem [{\citenamefont {Paris}\ \emph
  {et~al.}(2007{\natexlab{b}})\citenamefont {Paris}, \citenamefont {Baldwin},\
  and\ \citenamefont {Scalettar}}]{Paris07}%
  \BibitemOpen
  \bibfield  {author} {\bibinfo {author} {\bibfnamefont {N.}~\bibnamefont
  {Paris}}, \bibinfo {author} {\bibfnamefont {A.}~\bibnamefont {Baldwin}}, \
  and\ \bibinfo {author} {\bibfnamefont {R.~T.}\ \bibnamefont {Scalettar}},\
  }\href {\doibase 10.1103/PhysRevB.75.165113} {\bibfield  {journal} {\bibinfo
  {journal} {Phys. Rev. B}\ }\textbf {\bibinfo {volume} {75}},\ \bibinfo
  {pages} {165113} (\bibinfo {year} {2007}{\natexlab{b}})}\BibitemShut
  {NoStop}%
\bibitem [{\citenamefont {Castro~Neto}\ \emph {et~al.}(2009)\citenamefont
  {Castro~Neto}, \citenamefont {Guinea}, \citenamefont {Peres}, \citenamefont
  {Novoselov},\ and\ \citenamefont {Geim}}]{Castro09}%
  \BibitemOpen
  \bibfield  {author} {\bibinfo {author} {\bibfnamefont {A.~H.}\ \bibnamefont
  {Castro~Neto}}, \bibinfo {author} {\bibfnamefont {F.}~\bibnamefont {Guinea}},
  \bibinfo {author} {\bibfnamefont {N.~M.~R.}\ \bibnamefont {Peres}}, \bibinfo
  {author} {\bibfnamefont {K.~S.}\ \bibnamefont {Novoselov}}, \ and\ \bibinfo
  {author} {\bibfnamefont {A.~K.}\ \bibnamefont {Geim}},\ }\href {\doibase
  10.1103/RevModPhys.81.109} {\bibfield  {journal} {\bibinfo  {journal} {Rev.
  Mod. Phys.}\ }\textbf {\bibinfo {volume} {81}},\ \bibinfo {pages} {109}
  (\bibinfo {year} {2009})}\BibitemShut {NoStop}%
\bibitem [{\citenamefont {Kotov}\ \emph {et~al.}(2012)\citenamefont {Kotov},
  \citenamefont {Uchoa}, \citenamefont {Pereira}, \citenamefont {Guinea},\ and\
  \citenamefont {Castro~Neto}}]{Kotov12}%
  \BibitemOpen
  \bibfield  {author} {\bibinfo {author} {\bibfnamefont {V.~N.}\ \bibnamefont
  {Kotov}}, \bibinfo {author} {\bibfnamefont {B.}~\bibnamefont {Uchoa}},
  \bibinfo {author} {\bibfnamefont {V.~M.}\ \bibnamefont {Pereira}}, \bibinfo
  {author} {\bibfnamefont {F.}~\bibnamefont {Guinea}}, \ and\ \bibinfo {author}
  {\bibfnamefont {A.~H.}\ \bibnamefont {Castro~Neto}},\ }\href {\doibase
  10.1103/RevModPhys.84.1067} {\bibfield  {journal} {\bibinfo  {journal} {Rev.
  Mod. Phys.}\ }\textbf {\bibinfo {volume} {84}},\ \bibinfo {pages} {1067}
  (\bibinfo {year} {2012})}\BibitemShut {NoStop}%
\bibitem [{\citenamefont {Guzm\'an-Verri}\ and\ \citenamefont {Lew
  Yan~Voon}(2007)}]{Verri07}%
  \BibitemOpen
  \bibfield  {author} {\bibinfo {author} {\bibfnamefont {G.~G.}\ \bibnamefont
  {Guzm\'an-Verri}}\ and\ \bibinfo {author} {\bibfnamefont {L.~C.}\
  \bibnamefont {Lew Yan~Voon}},\ }\href {\doibase 10.1103/PhysRevB.76.075131}
  {\bibfield  {journal} {\bibinfo  {journal} {Phys. Rev. B}\ }\textbf {\bibinfo
  {volume} {76}},\ \bibinfo {pages} {075131} (\bibinfo {year}
  {2007})}\BibitemShut {NoStop}%
\bibitem [{\citenamefont {Vogt}\ \emph {et~al.}(2012)\citenamefont {Vogt},
  \citenamefont {De~Padova}, \citenamefont {Quaresima}, \citenamefont {Avila},
  \citenamefont {Frantzeskakis}, \citenamefont {Asensio}, \citenamefont
  {Resta}, \citenamefont {Ealet},\ and\ \citenamefont {Le~Lay}}]{Vogt12}%
  \BibitemOpen
  \bibfield  {author} {\bibinfo {author} {\bibfnamefont {P.}~\bibnamefont
  {Vogt}}, \bibinfo {author} {\bibfnamefont {P.}~\bibnamefont {De~Padova}},
  \bibinfo {author} {\bibfnamefont {C.}~\bibnamefont {Quaresima}}, \bibinfo
  {author} {\bibfnamefont {J.}~\bibnamefont {Avila}}, \bibinfo {author}
  {\bibfnamefont {E.}~\bibnamefont {Frantzeskakis}}, \bibinfo {author}
  {\bibfnamefont {M.~C.}\ \bibnamefont {Asensio}}, \bibinfo {author}
  {\bibfnamefont {A.}~\bibnamefont {Resta}}, \bibinfo {author} {\bibfnamefont
  {B.}~\bibnamefont {Ealet}}, \ and\ \bibinfo {author} {\bibfnamefont
  {G.}~\bibnamefont {Le~Lay}},\ }\href {\doibase
  10.1103/PhysRevLett.108.155501} {\bibfield  {journal} {\bibinfo  {journal}
  {Phys. Rev. Lett.}\ }\textbf {\bibinfo {volume} {108}},\ \bibinfo {pages}
  {155501} (\bibinfo {year} {2012})}\BibitemShut {NoStop}%
\bibitem [{\citenamefont {Houssa}\ \emph {et~al.}(2015)\citenamefont {Houssa},
  \citenamefont {Dimoulas},\ and\ \citenamefont {Molle}}]{Houssa15}%
  \BibitemOpen
  \bibfield  {author} {\bibinfo {author} {\bibfnamefont {M.}~\bibnamefont
  {Houssa}}, \bibinfo {author} {\bibfnamefont {A.}~\bibnamefont {Dimoulas}}, \
  and\ \bibinfo {author} {\bibfnamefont {A.}~\bibnamefont {Molle}},\ }\href
  {http://stacks.iop.org/0953-8984/27/i=25/a=253002} {\bibfield  {journal}
  {\bibinfo  {journal} {J. Phys.: Condens. Matter}\ }\textbf {\bibinfo {volume}
  {27}},\ \bibinfo {pages} {253002} (\bibinfo {year} {2015})}\BibitemShut
  {NoStop}%
\bibitem [{\citenamefont {Baskaran}(2018)}]{BaskaranSilicene}%
  \BibitemOpen
  \bibfield  {author} {\bibinfo {author} {\bibfnamefont {G.}~\bibnamefont
  {Baskaran}},\ }in\ \href {\doibase 10.1007/978-3-319-72374-7_5} {\emph
  {\bibinfo {booktitle} {Many-body Approaches at Different Scales}}},\ \bibinfo
  {editor} {edited by\ \bibinfo {editor} {\bibfnamefont {G.}~\bibnamefont
  {Angilella}}\ and\ \bibinfo {editor} {\bibfnamefont {C.}~\bibnamefont
  {Amovilli}}}\ (\bibinfo  {publisher} {Springer},\ \bibinfo {address}
  {Berlin},\ \bibinfo {year} {2018})\BibitemShut {NoStop}%
\bibitem [{\citenamefont {Raghu}\ \emph {et~al.}(2008)\citenamefont {Raghu},
  \citenamefont {Qi}, \citenamefont {Honerkamp},\ and\ \citenamefont
  {Zhang}}]{Raghu08}%
  \BibitemOpen
  \bibfield  {author} {\bibinfo {author} {\bibfnamefont {S.}~\bibnamefont
  {Raghu}}, \bibinfo {author} {\bibfnamefont {X.-L.}\ \bibnamefont {Qi}},
  \bibinfo {author} {\bibfnamefont {C.}~\bibnamefont {Honerkamp}}, \ and\
  \bibinfo {author} {\bibfnamefont {S.-C.}\ \bibnamefont {Zhang}},\ }\href
  {\doibase 10.1103/PhysRevLett.100.156401} {\bibfield  {journal} {\bibinfo
  {journal} {Phys. Rev. Lett.}\ }\textbf {\bibinfo {volume} {100}},\ \bibinfo
  {pages} {156401} (\bibinfo {year} {2008})}\BibitemShut {NoStop}%
\bibitem [{\citenamefont {Pairault}\ \emph {et~al.}(1998)\citenamefont
  {Pairault}, \citenamefont {S\'en\'echal},\ and\ \citenamefont
  {Tremblay}}]{Senechal98}%
  \BibitemOpen
  \bibfield  {author} {\bibinfo {author} {\bibfnamefont {S.}~\bibnamefont
  {Pairault}}, \bibinfo {author} {\bibfnamefont {D.}~\bibnamefont
  {S\'en\'echal}}, \ and\ \bibinfo {author} {\bibfnamefont {A.-M.~S.}\
  \bibnamefont {Tremblay}},\ }\href {\doibase 10.1103/PhysRevLett.80.5389}
  {\bibfield  {journal} {\bibinfo  {journal} {Phys. Rev. Lett.}\ }\textbf
  {\bibinfo {volume} {80}},\ \bibinfo {pages} {5389} (\bibinfo {year}
  {1998})}\BibitemShut {NoStop}%
\bibitem [{\citenamefont {Pairault}\ \emph {et~al.}(2000)\citenamefont
  {Pairault}, \citenamefont {S{\'{e}}n{\'{e}}chal},\ and\ \citenamefont
  {Tremblay}}]{Senechal2000}%
  \BibitemOpen
  \bibfield  {author} {\bibinfo {author} {\bibfnamefont {S.}~\bibnamefont
  {Pairault}}, \bibinfo {author} {\bibfnamefont {D.}~\bibnamefont
  {S{\'{e}}n{\'{e}}chal}}, \ and\ \bibinfo {author} {\bibfnamefont {A.-M.~S.}\
  \bibnamefont {Tremblay}},\ }\href {\doibase 10.1007/s100510070253} {\bibfield
   {journal} {\bibinfo  {journal} {Eur. Phys. J. B}\ }\textbf {\bibinfo
  {volume} {16}},\ \bibinfo {pages} {85} (\bibinfo {year} {2000})}\BibitemShut
  {NoStop}%
\bibitem [{\citenamefont {Wei\ss{}e}\ \emph {et~al.}(2006)\citenamefont
  {Wei\ss{}e}, \citenamefont {Wellein}, \citenamefont {Alvermann},\ and\
  \citenamefont {Fehske}}]{KPM}%
  \BibitemOpen
  \bibfield  {author} {\bibinfo {author} {\bibfnamefont {A.}~\bibnamefont
  {Wei\ss{}e}}, \bibinfo {author} {\bibfnamefont {G.}~\bibnamefont {Wellein}},
  \bibinfo {author} {\bibfnamefont {A.}~\bibnamefont {Alvermann}}, \ and\
  \bibinfo {author} {\bibfnamefont {H.}~\bibnamefont {Fehske}},\ }\href
  {\doibase 10.1103/RevModPhys.78.275} {\bibfield  {journal} {\bibinfo
  {journal} {Rev. Mod. Phys.}\ }\textbf {\bibinfo {volume} {78}},\ \bibinfo
  {pages} {275} (\bibinfo {year} {2006})}\BibitemShut {NoStop}%
\bibitem [{\citenamefont {Sota}\ and\ \citenamefont {Ito}(2007)}]{SotaRKPM}%
  \BibitemOpen
  \bibfield  {author} {\bibinfo {author} {\bibfnamefont {S.}~\bibnamefont
  {Sota}}\ and\ \bibinfo {author} {\bibfnamefont {M.}~\bibnamefont {Ito}},\
  }\href {\doibase 10.1143/JPSJ.76.054004} {\bibfield  {journal} {\bibinfo
  {journal} {J. Phys. Soc. Jpn.}\ }\textbf {\bibinfo {volume} {76}},\ \bibinfo
  {pages} {054004} (\bibinfo {year} {2007})}\BibitemShut {NoStop}%
\bibitem [{\citenamefont {Habibi}\ and\ \citenamefont
  {Jafari}(2013)}]{Habibi13}%
  \BibitemOpen
  \bibfield  {author} {\bibinfo {author} {\bibfnamefont {A.}~\bibnamefont
  {Habibi}}\ and\ \bibinfo {author} {\bibfnamefont {S.~A.}\ \bibnamefont
  {Jafari}},\ }\href {\doibase 10.1088/0953-8984/25/37/375501} {\bibfield
  {journal} {\bibinfo  {journal} {J. Phys.: Condens. Matter}\ }\textbf
  {\bibinfo {volume} {25}},\ \bibinfo {pages} {375501} (\bibinfo {year}
  {2013})}\BibitemShut {NoStop}%
\bibitem [{\citenamefont {Haase}\ \emph {et~al.}(2017)\citenamefont {Haase},
  \citenamefont {Yang}, \citenamefont {Pruschke}, \citenamefont {Moreno},\ and\
  \citenamefont {Jarrell}}]{Haase17}%
  \BibitemOpen
  \bibfield  {author} {\bibinfo {author} {\bibfnamefont {P.}~\bibnamefont
  {Haase}}, \bibinfo {author} {\bibfnamefont {S.-X.}\ \bibnamefont {Yang}},
  \bibinfo {author} {\bibfnamefont {T.}~\bibnamefont {Pruschke}}, \bibinfo
  {author} {\bibfnamefont {J.}~\bibnamefont {Moreno}}, \ and\ \bibinfo {author}
  {\bibfnamefont {M.}~\bibnamefont {Jarrell}},\ }\href {\doibase
  10.1103/PhysRevB.95.045130} {\bibfield  {journal} {\bibinfo  {journal} {Phys.
  Rev. B}\ }\textbf {\bibinfo {volume} {95}},\ \bibinfo {pages} {045130}
  (\bibinfo {year} {2017})}\BibitemShut {NoStop}%
\bibitem [{\citenamefont {Byczuk}\ \emph {et~al.}(2005)\citenamefont {Byczuk},
  \citenamefont {Hofstetter},\ and\ \citenamefont {Vollhardt}}]{Byczuk05}%
  \BibitemOpen
  \bibfield  {author} {\bibinfo {author} {\bibfnamefont {K.}~\bibnamefont
  {Byczuk}}, \bibinfo {author} {\bibfnamefont {W.}~\bibnamefont {Hofstetter}},
  \ and\ \bibinfo {author} {\bibfnamefont {D.}~\bibnamefont {Vollhardt}},\
  }\href {\doibase 10.1103/PhysRevLett.94.056404} {\bibfield  {journal}
  {\bibinfo  {journal} {Phys. Rev. Lett.}\ }\textbf {\bibinfo {volume} {94}},\
  \bibinfo {pages} {056404} (\bibinfo {year} {2005})}\BibitemShut {NoStop}%
\bibitem [{\citenamefont {Hafez}\ \emph {et~al.}()\citenamefont {Hafez},
  \citenamefont {Jafari},\ and\ \citenamefont {Abolhassani}}]{Hafez2009CUT}%
  \BibitemOpen
  \bibfield  {author} {\bibinfo {author} {\bibfnamefont {M.}~\bibnamefont
  {Hafez}}, \bibinfo {author} {\bibfnamefont {S.~A.}\ \bibnamefont {Jafari}}, \
  and\ \bibinfo {author} {\bibfnamefont {M.~R.}\ \bibnamefont {Abolhassani}},\
  }\href {\doibase 10.1016/j.physleta.2009.09.071} {\bibfield  {journal}
  {\bibinfo  {journal} {Phys. Lett. A}\ }\textbf {\bibinfo {volume} {373}},\
  \bibinfo {pages} {4479}}\BibitemShut {NoStop}%
\bibitem [{\citenamefont {Hafez}\ \emph {et~al.}(2010)\citenamefont {Hafez},
  \citenamefont {Jafari}, \citenamefont {Adibi},\ and\ \citenamefont
  {Shahbazi}}]{ShahnazHafez}%
  \BibitemOpen
  \bibfield  {author} {\bibinfo {author} {\bibfnamefont {M.}~\bibnamefont
  {Hafez}}, \bibinfo {author} {\bibfnamefont {S.~A.}\ \bibnamefont {Jafari}},
  \bibinfo {author} {\bibfnamefont {S.}~\bibnamefont {Adibi}}, \ and\ \bibinfo
  {author} {\bibfnamefont {F.}~\bibnamefont {Shahbazi}},\ }\href {\doibase
  10.1103/PhysRevB.81.245131} {\bibfield  {journal} {\bibinfo  {journal} {Phys.
  Rev. B}\ }\textbf {\bibinfo {volume} {81}},\ \bibinfo {pages} {245131}
  (\bibinfo {year} {2010})}\BibitemShut {NoStop}%
\bibitem [{\citenamefont {Sahebsara}\ and\ \citenamefont
  {S\'en\'echal}(2008)}]{Sahebsara08}%
  \BibitemOpen
  \bibfield  {author} {\bibinfo {author} {\bibfnamefont {P.}~\bibnamefont
  {Sahebsara}}\ and\ \bibinfo {author} {\bibfnamefont {D.}~\bibnamefont
  {S\'en\'echal}},\ }\href {\doibase 10.1103/PhysRevLett.100.136402} {\bibfield
   {journal} {\bibinfo  {journal} {Phys. Rev. Lett.}\ }\textbf {\bibinfo
  {volume} {100}},\ \bibinfo {pages} {136402} (\bibinfo {year}
  {2008})}\BibitemShut {NoStop}%
\bibitem [{\citenamefont {Amini}\ \emph {et~al.}(2009)\citenamefont {Amini},
  \citenamefont {Jafari},\ and\ \citenamefont {Shahbazi}}]{Amini09}%
  \BibitemOpen
  \bibfield  {author} {\bibinfo {author} {\bibfnamefont {M.}~\bibnamefont
  {Amini}}, \bibinfo {author} {\bibfnamefont {S.~A.}\ \bibnamefont {Jafari}}, \
  and\ \bibinfo {author} {\bibfnamefont {F.}~\bibnamefont {Shahbazi}},\ }\href
  {\doibase 10.1209/0295-5075/87/37002} {\bibfield  {journal} {\bibinfo
  {journal} {Eur. Phys. Lett.}\ }\textbf {\bibinfo {volume} {87}},\ \bibinfo
  {pages} {37002} (\bibinfo {year} {2009})}\BibitemShut {NoStop}%
\bibitem [{\citenamefont {Tikhonenko}\ \emph {et~al.}(2009)\citenamefont
  {Tikhonenko}, \citenamefont {Kozikov}, \citenamefont {Savchenko},\ and\
  \citenamefont {Gorbachev}}]{Gorbachevprl2009}%
  \BibitemOpen
  \bibfield  {author} {\bibinfo {author} {\bibfnamefont {F.~V.}\ \bibnamefont
  {Tikhonenko}}, \bibinfo {author} {\bibfnamefont {A.~A.}\ \bibnamefont
  {Kozikov}}, \bibinfo {author} {\bibfnamefont {A.~K.}\ \bibnamefont
  {Savchenko}}, \ and\ \bibinfo {author} {\bibfnamefont {R.~V.}\ \bibnamefont
  {Gorbachev}},\ }\href {\doibase 10.1103/PhysRevLett.103.226801} {\bibfield
  {journal} {\bibinfo  {journal} {Phys. Rev. Lett.}\ }\textbf {\bibinfo
  {volume} {103}},\ \bibinfo {pages} {226801} (\bibinfo {year}
  {2009})}\BibitemShut {NoStop}%
\bibitem [{\citenamefont {Garc\'{\i}a}\ \emph {et~al.}(2014)\citenamefont
  {Garc\'{\i}a}, \citenamefont {Uchoa}, \citenamefont {Covaci},\ and\
  \citenamefont {Rappoport}}]{Tatiana2014}%
  \BibitemOpen
  \bibfield  {author} {\bibinfo {author} {\bibfnamefont {J.~H.}\ \bibnamefont
  {Garc\'{\i}a}}, \bibinfo {author} {\bibfnamefont {B.}~\bibnamefont {Uchoa}},
  \bibinfo {author} {\bibfnamefont {L.}~\bibnamefont {Covaci}}, \ and\ \bibinfo
  {author} {\bibfnamefont {T.~G.}\ \bibnamefont {Rappoport}},\ }\href {\doibase
  10.1103/PhysRevB.90.085425} {\bibfield  {journal} {\bibinfo  {journal} {Phys.
  Rev. B}\ }\textbf {\bibinfo {volume} {90}},\ \bibinfo {pages} {085425}
  (\bibinfo {year} {2014})}\BibitemShut {NoStop}%
\bibitem [{\citenamefont {Habibi}\ \emph
  {et~al.}(2018{\natexlab{b}})\citenamefont {Habibi}, \citenamefont {Jafari},\
  and\ \citenamefont {Rouhani}}]{Habibi2018Resilience}%
  \BibitemOpen
  \bibfield  {author} {\bibinfo {author} {\bibfnamefont {A.}~\bibnamefont
  {Habibi}}, \bibinfo {author} {\bibfnamefont {S.~A.}\ \bibnamefont {Jafari}},
  \ and\ \bibinfo {author} {\bibfnamefont {S.}~\bibnamefont {Rouhani}},\ }\href
  {\doibase 10.1103/PhysRevB.98.035142} {\bibfield  {journal} {\bibinfo
  {journal} {Phys. Rev. B}\ }\textbf {\bibinfo {volume} {98}},\ \bibinfo
  {pages} {035142} (\bibinfo {year} {2018}{\natexlab{b}})}\BibitemShut
  {NoStop}%
\bibitem [{\citenamefont {Oseledec}(1968)}]{Oseledec}%
  \BibitemOpen
  \bibfield  {author} {\bibinfo {author} {\bibfnamefont {V.~I.}\ \bibnamefont
  {Oseledec}},\ }\href {https://ci.nii.ac.jp/naid/10004591080/en/} {\bibfield
  {journal} {\bibinfo  {journal} {Trans.Moscow Math.Soc.}\ }\textbf {\bibinfo
  {volume} {19}},\ \bibinfo {pages} {197} (\bibinfo {year} {1968})}\BibitemShut
  {NoStop}%
\bibitem [{\citenamefont {MacKinnon}\ and\ \citenamefont
  {Kramer}(1981)}]{MacKinnon81}%
  \BibitemOpen
  \bibfield  {author} {\bibinfo {author} {\bibfnamefont {A.}~\bibnamefont
  {MacKinnon}}\ and\ \bibinfo {author} {\bibfnamefont {B.}~\bibnamefont
  {Kramer}},\ }\href {\doibase 10.1103/PhysRevLett.47.1546} {\bibfield
  {journal} {\bibinfo  {journal} {Phys. Rev. Lett.}\ }\textbf {\bibinfo
  {volume} {47}},\ \bibinfo {pages} {1546} (\bibinfo {year}
  {1981})}\BibitemShut {NoStop}%
\bibitem [{\citenamefont {MacKinnon}\ and\ \citenamefont
  {Kramer}(1983)}]{MacKinnon83}%
  \BibitemOpen
  \bibfield  {author} {\bibinfo {author} {\bibfnamefont {A.}~\bibnamefont
  {MacKinnon}}\ and\ \bibinfo {author} {\bibfnamefont {B.}~\bibnamefont
  {Kramer}},\ }\href {\doibase 10.1007/bf01578242} {\bibfield  {journal}
  {\bibinfo  {journal} {Zeitschrift für Physik B Condensed Matter}\ }\textbf
  {\bibinfo {volume} {53}},\ \bibinfo {pages} {1} (\bibinfo {year}
  {1983})}\BibitemShut {NoStop}%
\bibitem [{\citenamefont {Habibi}\ \emph
  {et~al.}(2018{\natexlab{c}})\citenamefont {Habibi}, \citenamefont {Ghadami},
  \citenamefont {Jafari},\ and\ \citenamefont
  {Rouhani}}]{habibi2018topological}%
  \BibitemOpen
  \bibfield  {author} {\bibinfo {author} {\bibfnamefont {A.}~\bibnamefont
  {Habibi}}, \bibinfo {author} {\bibfnamefont {R.}~\bibnamefont {Ghadami}},
  \bibinfo {author} {\bibfnamefont {S.~A.}\ \bibnamefont {Jafari}}, \ and\
  \bibinfo {author} {\bibfnamefont {S.}~\bibnamefont {Rouhani}},\ }\href
  {https://arxiv.org/abs/1807.01339} {\bibfield  {journal} {\bibinfo  {journal}
  {arXiv preprint arXiv:1807.01339}\ } (\bibinfo {year}
  {2018}{\natexlab{c}})}\BibitemShut {NoStop}%
\end{thebibliography}%
\end{document}